\documentclass{article}
\usepackage{xcolor}
\usepackage{times}
\usepackage{amssymb}
\usepackage{amsmath,amsthm}
\usepackage{graphicx}
\usepackage{multirow}
\usepackage[authoryear]{natbib}
\usepackage{tabularx}
\usepackage{mathtools}
\usepackage{enumitem}
\usepackage{commath}
\usepackage[affil-it]{authblk} 
\usepackage{url}
\usepackage[hmargin=1.25in,vmargin=1.15in]{geometry}

\newcommand{\bbE}{\mathbb{E}}
\DeclareMathOperator{\diag}{\text{diag}}

\DeclareMathOperator{\trace}{tr}

\newcommand{\bbR}{\mathbb{R}}

\newcommand{\cN}{\mathcal{N}}
\newcommand{\cI}{\mathcal{I}}
\newcommand{\cO}{\mathcal{O}}
\newcommand{\cL}{\mathcal{L}}
\newcommand{\cX}{\mathcal{X}}
\newcommand{\cY}{\mathcal{Y}}

\newcommand{\mui}{\mu_{\infty}}

\newcommand{\scal}[2]{\langle{#1},{#2}\rangle}
\newcommand{\nor}[1]{\|{#1}\|}

\newcommand{\tr}{\top\!}

\newcommand{\lmin}{\underline{\lambda}}
\newcommand{\dB}{d\mathbf{B}}
\newcommand{\bW}{\mathbf{W}}
\newcommand{\bB}{\mathbf{B}}
\newcommand{\bw}{\mathbf{w}}
\newcommand{\bc}{\mathbf{c}}
\newcommand{\bu}{\mathbf{u}}
\newcommand{\bue}{\mathbf{u}_{\varepsilon}}
\newcommand{\bx}{\mathbf{x}}
\newcommand{\by}{\mathbf{y}}
\newcommand{\bZ}{\mathbf{Z}}
\newcommand{\bN}{\mathbf{N}}
\newcommand{\bU}{\mathbf{U}}

\newcommand{\bbone}{\mathbf{1}}
\newcommand{\bmu}{\boldsymbol{\mu}}

\newtheorem{theorem}{Theorem}[section]

\newtheorem{proposition}{Proposition}[section]

\theoremstyle{definition}

\author{{\large Jake Bouvrie}}
\affil{Laboratory for Computational and Statistical Learning, Massachusetts Institute of Technology, 
Cambridge, MA USA and Instituto Italiano di Tecnologia, Genova, Italy \\
\texttt{jvb@csail.mit.edu}}
\author{{\large Jean-Jacques Slotine}}
\affil{Nonlinear Systems Laboratory,
Massachusetts Institute of Technology,
Cambridge, MA USA\\
\texttt{jjs@mit.edu}}

\title{Synchronization and Noise: A Mechanism for Regularization in Neural Systems}

\begin{document}
\maketitle
%

\begin{abstract}
To learn and reason in the presence of uncertainty, the brain must be capable of imposing some form of regularization. Here we suggest, through theoretical and computational arguments, that the combination of noise with synchronization provides a plausible mechanism for regularization in the nervous system. The functional role of regularization is considered in a general context in which coupled computational systems receive inputs corrupted by correlated noise. Noise on the inputs is shown to impose regularization, and when synchronization upstream induces time-varying correlations across noise variables, the degree of regularization can be calibrated over time. The proposed mechanism is explored first in the context of a simple associative learning problem, and then in the context of a hierarchical sensory coding task. The resulting qualitative behavior coincides with experimental data from visual cortex.
\end{abstract}

\section{Introduction}
The problem of learning from examples is in most circumstances ill-posed. This is particularly true for biological organisms, where the ``examples'' are often complex and few in number, and the ability to adapt is a matter of survival. Theoretical work in inverse problems has long established that regularization restores well-posedness~\citep{EnglBook,PoggioSmale:AMS:03} and furthermore, implies stability and generalization of a learned rule~\citep{Bousquet:JMLR:02}. How the nervous system imposes regularization is not entirely clear, however. Bayesian theories of learning and decision making~\citep{Kording:Review:06,Shadlen:Science:09,Shadlen:Nature:07,RoyLlinasChapter:12} hold that that brain is able to represent prior distributions and assign (time-varying) uncertainty to sensory measurements. By way of a Bayesian integration, the brain may effectively work with hypothesis spaces of limited complexity when appropriate, trading off prior knowledge against new evidence~\citep{Shadlen:JN:11}. But while these mechanisms can effect regularization, it is still not clear how to calibrate it: when to cease adaptation or how to fix a hypothesis space suited to a given task. A second possible explanation is that regularization -- and a representation of uncertainty -- may emerge naturally due to noise. Intuitively, if noise is allowed to ``smear'' observations presented to a learning apparatus, overfitting may be mitigated -- a well known phenomenon in artificial neural networks~\citep{Bishop:NECO:95}. In this sense, noise may be more abstractly interpreted, or even {\em defined}, as ``a family of hypotheses about the possible forms of knowledge'', to quote from~\citep{RoyLlinasChapter:12}.

In this paper we argue that noise provides an appealing, plausible mechanism for regularization in the nervous system. We consider a general context in which coupled computational circuits subject to independent noise receive common inputs corrupted by spatially correlated noise.  Information processing pathways in the mammalian visual cortex, for instance, fall under such an organizational pattern~\citep{Lisberger:09,Smith:JN:08,Gawne:JN:93}. The computational systems in this setting represent high-level processing stages, downstream from localized populations of neurons which encode sensory input. Noise correlations in the latter arise from, for instance, within-population recurrent connections, shared feed-forward inputs, and common stimulus preferences~\citep{Smith:JN:08}. Independent noise impacting higher-level computational elements may arise from more intrinsic, ambient neuronal noise sources, and may be roughly independent due to broader spatial distribution~\citep{Faisal:NRN:2008}.


To help understand the functional role of noise in inducing regularization, we propose a high-level model that can explain quantitatively how noise translates into regularization, and how regularization may be calibrated over time. The ability to adjust regularization is key: as an organism accumulates experience, its models of the world should be able to adjust to the complexity of the relationships and phenomena it encounters, as well as reconcile new information with prior probabilities. 
Our point of view is complementary to Bayesian theories of learning; the representation and integration of sensory uncertainty is closely related to a regularization interpretation of learning in ill-posed settings. 

We suggest that regularization may be plausibly controlled by one of the most ubiquitous mechanisms in the brain: synchronization. A simple, one-dimensional regression (association) problem in the presence of both independent ambient noise and correlated measurement noise suffices to illustrate the core ideas (Sections~\ref{sec:gradient-learning} and~\ref{sec:noise-gives-reg}). When a learner is presented with a collection of noisy observations, we show that synchronization may be used to adjust the dependence between observational noise variables, and that this in turn leads to a quantifiable change in the degree of regularization imposed upon the learning task (Section~\ref{sec:calibrating-reg}). Here, regularization is further shown to both improve the convergence rate towards the solution to the regression problem, {\em and} reduce the negative impact of ambient noise. We then consider a more complex instance of noise-induced regularization, in which a hierarchical {\em predictive coding network} of the type introduced by~\cite{Rao:NNS:99} encodes a visual stimulus in terms of an arbitrary but fixed pair of dictionaries (Section~\ref{sec:learning-rao-ballard}). In particular, a linear, two-layer network in continuous-time with additive noise on the top-down and bottom-up error signals, as well as on the dictionary itself, is explored. This particular hierarchy is shown to follow a Langevin dynamics similar to that of the simple regression problem, so that some of the key insights relating to the regression system readily apply. We show that noise on the encoding dictionaries has the effect of regularizing the underlying optimization problem of interest, and it is revealed that dictionary noise  interacts with noise on the error signals (in the role of ambient noise) in a manner which, unlike in the regression setting, may not always mitigate the effect of error signal noise.

The qualitative behavior of these models can be adjusted to coincide with experimental data from visual tracking tasks~\citep{Lisberger:09} (area MT) and from anesthetized animals~\citep{Smith:JN:08} (area V1), in which correlated noise impacts sensory measurements and  correlations increase over short time scales. Other experiments involving perceptual learning tasks have shown that noise correlations decrease with long-term training~\citep{Gu:Neuron:11}, and this behavior may also be captured by the model. The mechanism we propose suggests that changes in noise correlations arising from feedback synchronization can calibrate regularization, possibly leading to improved convergence properties or better solutions. Collectively, the experimental evidence lends credence to the hypothesis that, at a high level, the brain may be optimizing its learning processes by adapting dependence among noise variables, with regularization an underlying computational theme. Simulations demonstrating these ideas are presented in Section~\ref{sec:simulations}.

Lastly, we consider how continuous dynamics solving a given learning problem might be efficiently computed in cortex (Section~\ref{sec:distrib-computation}; see Section~\ref{sec:simulations} for simulations). In addition to supporting regularization, noise can be harnessed to facilitate distributed computation of the gradients needed to implement a dynamic optimization process. Following from this observation, we analyze a continuous-time, stochastic finite difference scheme approximating derivatives of quadratic objectives. Difference signals and approximately independent perturbations are the only required computational components. This distributed approach to the implementation of dynamic learning processes further highlights a connection between parallel stochastic gradient descent algorithms~\citep{Spall:TAC:92,JabriFlower:92,Vorontsov:97}, and neural computation.

Some of the material in this paper was previously presented in an abbreviated form at the Neural Information Processing Systems Conference in 2012; see~\citep{Bouvrie:NIPS:12}.

\section{Learning as noisy gradient descent on a network}
\label{sec:gradient-learning}
The learning process we will consider is that of a one-dimensional linear fitting problem described by a dynamic gradient based minimization of a square loss objective, in the spirit of Rao \& Ballard~\citep{Rao:NNS:99}. This is perhaps the simplest and most fundamental abstract learning problem that an organism might be confronted with -- that of using experiential evidence to infer correlations and ultimately discover causal relationships which govern the environment and which can be used to make predictions about the future. The model realizing this learning process is also simple, in that we capture neural communication as an abstract process ``in which a neural element (a single neuron or a population of neurons) conveys certain aspects of its functional state to another neural element''~\citep{Schnitzler:NatRev:05}. In doing so, we focus on the underlying computations taking place in the nervous system rather than particular neural representations. 
The analysis that follows, however, may be extended more generally to multi-layer feedback hierarchies.

To make the setting more concrete, assume that we have observed a set of input-output examples $\{x_i\in\bbR,y_i\in\bbR \}_{i=1}^m$, with each $x_i$ representing a generic unit of sensory experience, and want to estimate the linear regression function $f_w(x) = wx$ (we assume the intercept is 0 for simplicity). Adopting the square loss, the total prediction error incurred on the observations by the rule $f_w$ is given by 
\begin{equation}\label{eqn:obective_fn}
E(w) = \tfrac{1}{2}\sum_{i=1}^m(y_i - f_{w}(x_i))^2 = \tfrac{1}{2}\sum_{i=1}^m(y_i - wx_i)^2.
\end{equation}
Note that there is no explicit regularization penalty here.
We will model adaptation (training) by a noisy gradient descent process on this squared prediction error loss function. The gradient of $E$ with respect to the slope parameter is given by $\nabla_{w} E =  -\sum_{i=1}^m(y_i - wx_i)x_i$, and generates the continuous-time, noise-free gradient dynamics 
\begin{equation}\label{eqn:noise-free-lr}
\dot{w}=-\nabla_{w} E(w).
\end{equation}


The learning dynamics we will consider, however, are assumed to be corrupted by two distinct kinds of noise:
\begin{enumerate}\itemsep 0pt 
\item[(N1)] Sensory observations $(x_i)_i$ are corrupted by time-varying, correlated noise processes.
\item[(N2)] The dynamics are themselves corrupted by additive ``ambient'' noise. 
\end{enumerate}
To accommodate (N1) we will borrow an averaging or, {\em homogenization}, technique for multi-scale systems of stochastic differential equations (SDEs) that will drastically simplify analysis. 
We have discussed the origins of (N1) above. The noise (N2) may be significant (we do not take small noise limits) and can be attributed to some or all of: error in computing and sensing a gradient, intrinsic neuronal noise~\citep{Faisal:NRN:2008} (aggregated or localized), or interference between large assemblies of neurons or circuits.


Synchronization among circuits and/or populations will be modeled by considering multiple coupled dynamical systems, each receiving the same noisy observations. Such networks of systems capture common pooling or averaging computations, and provides a means for studying variance reduction.
The {\em collective enhancement of precision} hypothesis suggests that the nervous system copes with noise by averaging over collections of signals in order to reduce variation in behavior and improve computational accuracy~\citep{Sherman:BJ:1991,Kinard:BJ:91,Tabareau10,Bouvrie:NECO:11}.
Coupling synchronizes the collection of dynamical systems so that each tends to a common ``consensus'' trajectory having reduced variance. If the coupling is strong enough, then the variance of the consensus trajectory decreases as $\cO(1/n)$ after transients, if there are $n$ signals or circuits~\citep{Sherman:BJ:1991,Needleman:PD:2001,Pham09,Bouvrie:NECO:11}. For nonlinear systems, synchronization is essential for realizing this form of noise reduction; averaging over discordant trajectories will reduce variance, but will also destroy information content~\citep{Tabareau10}. As real neural circuits found in biology exhibit non-linear behavior (consider saturation effects), synchronization is compelling just from a noise elimination perspective. In the case of linear systems, although global averaging over uncoupled systems is possible, coupling linear systems still brings significant advantages: an improved signal may be read from any of the individual elements, and global pooling may be replaced by local intercommunication.

We will consider regularization in the context of networks of coupled SDEs, and investigate the impact of coupling, redundancy ($n$) and regularization upon the convergence behavior of the system. Considering networks will allow a more general analysis of the interplay between different mechanisms for coping with noise, however $n$ can be small or 1 in some situations.

Formally, the noise-free flow~\eqref{eqn:noise-free-lr} can be modified to include noise sources (N1) and (N2) as follows. Noise (N1) may be modeled as a white-noise limit of Ornstein-Uhlenbeck  (OU) processes $(\bZ_t)_i$, and (N2) as an additive diffusive noise term. In differential form, we have
\begin{subequations}\label{eqn:uncoupled-noise-system}
\begin{align}
dw_t &= -\bigl(w_t\|\bx+\bZ_t\|^2 - \scal{\bx + \bZ_t}{\by}\bigr)dt + \sigma dB_t \label{eqn:w-dyn}\\
dZ_t^i &= -\frac{Z_t^i}{\varepsilon}dt + \frac{\sqrt{2}\gamma}{\sqrt{\varepsilon}} dB_t^i,\qquad i=1,\ldots,m \label{eqn:ou-noise} .
\end{align}
\end{subequations}
Here, $B_t$ denotes the standard 1-dimensional Brownian motion and captures noise source (N2). The observations $(\bx)_i = x_i$ are corrupted by the noise processes $(\bZ_t)_i=Z^i_t$, following (N1). For the moment, the $Z^i_t$ are independent, but we will relax this assumption later. The parameter $0< \varepsilon \ll 1$ controls the correlation time of a given noise process. In the limit as  $\varepsilon\to 0$, $Z_t^i$ may be viewed as a family of independent zero-mean Gaussian random variables indexed by $t$. Characterizing the noise $\bZ_t$ as~\eqref{eqn:ou-noise} with $\varepsilon\to 0$ serves as both a modeling approximation/idealization and an analytical tool.

%

\subsection{Homogenization}
\label{sec:averaging-regression-system}
The system~\eqref{eqn:w-dyn}-\eqref{eqn:ou-noise} above is a classic ``fast-slow'' system: the gradient descent trajectory $w_t$ evolves on a timescale much longer than the $\cO(\varepsilon)$ stochastic perturbations $\bZ_t$. Homogenization considers the dynamics of $w_t$ after averaging out the effect of the fast variable $\bZ_t$. In the limit as $\varepsilon\to 0$ in~\eqref{eqn:ou-noise}, the solution to the averaged SDE converges (in a sense to be discussed below) to the solution of the original SDE~\eqref{eqn:w-dyn}.

The following Theorem is an instance of~\cite[Thm. 3]{PardVeretI}, adapted to the present setting.
\begin{theorem}\label{thm:averaging}
Let $0<\varepsilon\ll 1$, $\sigma,\gamma>0$ and let $\cX,\cY$ denote finite-dimensional Euclidean spaces. Consider the system
\begin{subequations}\label{eqn:homog-thm-system}
\begin{align}
dx &= f(x,y)dt + \gamma dW_t, & x(0) & = x_0 \label{eqn:dx_generic}\\
dy &= \varepsilon^{-1}g(y)dt + \varepsilon^{-1/2}\sigma dB_t, & y(0)&=y_0,
\label{eqn:dy_generic}
\end{align}
\end{subequations}
where $x\in\cX,y\in\cY$, and $W_t\in\cX, B_t\in\cY$ are independent multivariate Brownian motions. Assume that for all $x\in\cX, y\in\cY$ the following conditions on~\eqref{eqn:homog-thm-system} hold:
\begin{align*}
\scal{g(y)}{y/\nor{y}} &\leq -r\nor{y}^{\alpha}, \qquad \\
\nor{f(x,y)-f(x',y)} &\leq C(y)\nor{x-x'} \\
\nor{f(x,y)} &\leq K(1+\nor{x})(1+\nor{y}^q),
\end{align*}
with $r>0, \alpha \geq 0, q<\infty$, and where $C(y)$ is a constant depending on $y$.
 If the SDE~\eqref{eqn:dy_generic} is ergodic, then there exists a unique invariant measure $\mu_{\infty}$ characterizing the probability distribution of $y_t$ in the steady state, and we may define the vector field
$
F(x) \triangleq \bbE_{\mu_{\infty}}[f(x,y)] = \int_{\cY}f(x,y)\mu_{\infty}(dy) .
$
Furthermore, $x(t)$ solving~\eqref{eqn:dx_generic} is closely approximated by $X(t)$ solving
$$
dX = F(X)dt + \gamma dW_t, \quad X(0)=x_0
$$
in the sense that, for any $t\in[0,T]$, $x(t)\Rightarrow X(t)$ in $C([0,T],\cX)$ as $\varepsilon\to 0$.
\end{theorem}

It may be readily shown that the system~\eqref{eqn:uncoupled-noise-system} satisfies the conditions of Theorem~\ref{thm:averaging}. Moreover, the OU process~\eqref{eqn:ou-noise} on $\bbR^{m}$ is known to be ergodic with stationary distribution  $\bZ_{\infty}\sim\cN(\mathbf{0}, \gamma^2 I)$ (see e.g.~\citep{Kallenberg}), where $\cN(\mu,\Sigma)$ denotes the multivariate Gaussian distribution with mean $\mu$ and covariance $\Sigma$. Averaging over the fast variable $\bZ_t$ appearing in~\eqref{eqn:w-dyn} with respect to this distribution gives
\begin{equation}\label{eqn:w-avg}
dw_t = -\bigl[w_t(\nor{\bx}^2 + m\gamma^2) - \scal{\bx}{\by}\bigr]dt + \sigma dB_t,
\end{equation}
and by Theorem~\ref{thm:averaging}, we can conclude that Equation~\eqref{eqn:w-avg} well-approximates~\eqref{eqn:w-dyn} when $\varepsilon\to 0$ in~\eqref{eqn:ou-noise} in the sense of weak convergence of probability measures.

\subsection{Network structure}
Now consider $n\geq 1$ diffusively coupled neural systems implementing the dynamics~\eqref{eqn:w-avg}, with associated parameters $\bw(t)=\bigl(w_1(t),\ldots,w_n(t)\bigr)$. If $W_{ij}\geq 0$ is the coupling strength between systems $i$ and $j$, 
$L=\diag(W\bbone) - W$ is the {\em network Laplacian}~\citep{MesbahiBook}. We assume here that $L$ is symmetric and defines a connected network graph.
Letting $\alpha:=\nor{\bx}^2 + m\gamma^2$, $\beta:=\scal{\bx}{\by}$ and $\bmu:=(\beta/\alpha)\bbone$, the coupled system can be written concisely as
\begin{equation}\label{eqn:w-coupled}
\begin{aligned}
d\bw_t &= -(L+\alpha I)\bw_t dt + \beta\bbone dt + \sigma \dB_t\\
 &= (L+\alpha I)(\bmu - \bw_t)dt + \sigma\dB_t \,,
 \end{aligned}
\end{equation}
with $\bB_t$ an $n$-dimensional Brownian motion. The diffusive couplings here should be interpreted as modeling abstract intercommunication between and among different neural circuits, populations, or pathways. In such a general setting, diffusive coupling is a natural and mathematically tractable choice that can capture the key, aggregate aspects of communication among neural systems. 
Note that one can equivalently consider $n$ systems~\eqref{eqn:w-dyn} and then homogenize assuming $n$ copies of the same noise process $\bZ_t$, {\em or} $n$ independent noise processes $\{\bZ_t^{(i)}\}_i$; either choice also leads to~\eqref{eqn:w-coupled}.

\section{Learning with noisy data imposes regularization}
\label{sec:noise-gives-reg}
Equation~\eqref{eqn:w-coupled} is seen by inspection to be of Langevin type, and has as its solution (see e.g.~\citep{OksendalBook}) the OU process
\begin{equation}
\label{eqn:w-ou-soln}
\bw(t) = e^{-(L+\alpha I)t}\bw(0) + \bigl(I - e^{-(L+\alpha I)t}\bigr)\bmu  +
\sigma\int_0^t e^{-(L+\alpha I)(t-s)}\dB_s .
\end{equation}
Integrals of Brownian motion are normally distributed, so $\bw(t)$ is a Gaussian process and can be characterized entirely by its time-dependent mean and covariance, $\bw(t)\sim\cN\bigl(\bmu_w(t),\Sigma_w(t)\bigr)$. A straightforward manipulation gives 
\begin{align}
\bmu_w(t) :&=\bbE[\bw(t)] = e^{-(L+\alpha I)t}\bbE[\bw(0)] + \bigl(I - e^{-(L+\alpha I)t}\bigr)\bmu\label{eqn:w-mean-process}\\
\Sigma_w(t) :&= \bbE\left[\bigl(\bw(t) -\bbE\bw(t)\bigr)\bigl(\bw(t) -\bbE\bw(t)\bigr)^{\!\tr}\right]\notag\\
 &= e^{-(L+\alpha I)t}\bbE[\bw(0)\bw(0)^{\tr}]e^{-(L+\alpha I)t}
+ \frac{\sigma^2}{2}(L+\alpha I)^{-1}\bigl(I - e^{-2(L+\alpha I)t}\bigr) .\notag
\end{align}
The solution to the noise-free regression problem (minimizing~\eqref{eqn:obective_fn}) is given by $w^{*}=\scal{\bx}{\by}/\nor{\bx}^2$, however~\eqref{eqn:w-ou-soln} together with~\eqref{eqn:w-mean-process} reveals that, for any $i\in\{1,\ldots,n\}$,
\begin{equation}\label{eqn:reg_soln}
\bbE[w_i(t)]\xrightarrow{t\to\infty}(\bmu)_i=\frac{\scal{\bx}{\by}}{\nor{\bx}^2 + m\gamma^2}
\end{equation}
which is exactly the solution to the {\em regularized} regression problem
\begin{equation}
\label{eqn:regularized-opt-problem}
\min_{w\in\bbR}\|\by-w\bx\|^2 + \lambda w^2
\end{equation}
with regularization parameter $\lambda:=m\gamma^2$. To summarize, we have considered a network of coupled, noisy gradient flows implementing unregularized linear regression. When the observations $\bx$ are noisy, all elements of the network converge in expectation to a common equilibrium point representing a regularized solution to the original regression problem.


\subsection{Convergence behavior}
\label{sec:reg-sys-conv-behavior}
In the previous section we showed that the network converges to the solution of a regularized regression problem, but left open a few important questions: What determines the convergence rate? How does the noise (N1),(N2) impact convergence? How does coupling and redundancy (number of circuits $n$) impact convergence? How do these quantities affect the variance of the error? We can address these questions by decomposing $\bw(t)$ into orthogonal components, $\bw(t) = \bar{w}(t)\bbone + \widetilde{\bw}(t)$, representing the mean-field trajectory $\bar{w}=\tfrac{1}{n}\bbone^{\tr}\bw$, and fluctuations about the mean  $\widetilde{\bw}=\bw-\bar{w}\bbone$. We may then study the error
\begin{equation}\label{eqn:sys-err-decomp}
\bbE\bigl[\tfrac{1}{n}\nor{\bw(t) - \bmu}^2\bigr] =
\bbE\bigl[\tfrac{1}{n}\nor{\widetilde{\bw}(t)}^2\bigr] +
\bbE\bigl[\tfrac{1}{n}\nor{\bar{w}(t)\bbone - \bmu}^2\bigr]
\end{equation}
by studying each term separately. Decomposing the error into fluctuations about the average and the distance between the average and the noise-free equilibrium allows one to see that there are actually two different convergence rates governing the system: one determines convergence towards the synchronization subspace (where $\widetilde{\bw}=0$), and the another determines convergence to the equilibrium point $\bmu$. The following result provides quantitative answers to the questions posed above:
\begin{theorem}\label{thm:sys-convergence}
Let $\widetilde{C}, \overline{C}$ be constants which do not depend on time, and let $\lmin$ denote the smallest non-zero eigenvalue of $L$. Given a regularization parameter $\lambda > 0$ determining an optimization problem of the form~\eqref{eqn:regularized-opt-problem}, let $\alpha:=\nu(\nor{\bx}^2+\lambda)$ for some fixed $\nu>0$ and set $\bmu:=(\scal{\bx}{\by}\nu/\alpha)\bbone$. Then for all $t>0$, any solution to 
\[
d\bw_t = (L+\alpha I)(\bmu - \bw_t)dt + \sigma\dB_t
\]
 satisfies
\begin{equation}\label{eqn:full-error}
\bbE\bigl[\tfrac{1}{n}\nor{\bw(t) - \bmu}^2\bigr] \leq
 \widetilde{C}e^{-2(\lmin + \alpha)t} +
 \overline{C}e^{-2\alpha t} +
 \frac{\sigma^2}{2}\left(\frac{1}{\lmin+\alpha} + \frac{1}{\alpha n}\right).
\end{equation}
\end{theorem}
A proof is given in the Appendix. For the system described by Equation~\eqref{eqn:w-coupled} above, $\nu\equiv 1$,  and $\lambda=m\gamma^2$ so that $\alpha=\nor{\bx}^2 + m\gamma^2$. The first term of~\eqref{eqn:full-error} estimates the transient part of the fluctuations term in~\eqref{eqn:sys-err-decomp}, and we find that the rate of convergence to the synchronization subspace is $2(\lmin + \alpha)$. The second term term estimates the transient part of  the centroid's trajectory, and we see that the rate of convergence of the mean trajectory to equilibrium is $2\alpha$. 
In the presence of noise, however, the system will neither synchronize nor reach equilibrium exactly. After transients, we see that the residual error is given by the last term in~\eqref{eqn:full-error}. This term quantifies the steady-state interaction between: gradient noise ($\sigma$); regularization ($\alpha$, via the observation noise $\gamma$); network topology (via $\lmin$), coupling strength (via $\lmin$), and redundancy ($n$; possibly $\lmin$). 

\subsection{Discussion}\label{sec:reg-remarks}
From the results above we can draw a few conclusions about networks of noisy learning systems:
\begin{enumerate}[itemsep=0pt,leftmargin=0.45cm]
\item Regularization improves both the synchronization rate and the rate of convergence to equilibrium.
\item Regularization contributes towards {\em reducing} the effect of the gradient noise $\sigma$: (N1) counteracts (N2).
\item Regularization {\em changes the solution}, so we cannot view regularization as a ``free-parameter'' that can be used solely to improve convergence or reduce noise. Faster convergence rates and noise reduction should be viewed as beneficial side-effects, while the appropriate degree of regularization primarily depends on the learning problem at hand.
\item  The number of circuits $n$ and the coupling strength contribute towards reducing the effect of the gradient noise (N2) (that is, the variance of the error) and improve the synchronization rate, but do not affect the rate of convergence toward equilibrium.
\item Coupling strength and redundancy {\em cannot} be used to control the degree of regularization, since the equilibrium solution $\bmu$ does not depend on $n$ or the spectrum of $L$. This is true no matter how the coupling weights $W_{ij}$ are chosen, since constants will always be in the null space of $L$ and $\bmu$ is a constant vector.
\end{enumerate}
In the next section we will show that if the noise processes  $\{Z_t^i\}_i$ are {\em themselves} trajectories of a coupled network, then synchronization {\em can} be a mechanism for controlling the regularization imposed on a learning process. 



\section{Calibrating regularization with synchronization}
\label{sec:calibrating-reg}
If instead of assuming independent noise processes corrupting the data as in~\eqref{eqn:ou-noise}, we consider correlated noise variables $(Z^i_t)_{i=1}^m$, it is possible for synchronization to control the regularization which the noise imposes on a learning system of the form~\eqref{eqn:w-dyn}. 
A collection of dependent observational noise processes is perhaps most conveniently modeled by coupling the OU dynamics~\eqref{eqn:ou-noise} introduced before through another (symmetric) network Laplacian $L_z$:
\begin{equation}\label{eqn:noise-network}
d\bZ_t = -\frac{1}{\varepsilon}(L_z + \eta I)\bZ_t dt + \frac{\sqrt{2}\gamma}{\sqrt{\varepsilon}}\dB_t,
\end{equation}
for some $\eta>0$.
We now have two networks: the first network of gradient systems is the same as before, but the observational noise process $\bZ_t$ is now generated by another network. For purposes of analysis, this model suffices to capture generalized correlated noise sources. In the actual biology, however,  correlations may arise in a number of possible ways, which may or may not include diffusively coupled dynamic noise processes.

To analyze what happens when a network of learning systems~\eqref{eqn:w-dyn} is driven by observation noise of the form~\eqref{eqn:noise-network}, we take an approach similar to that of the previous Section. The first step is again homogenization. The system~\eqref{eqn:noise-network} may be viewed as a zero-mean variation of~\eqref{eqn:w-coupled}, and its solution $\bZ_t\sim\cN\bigl(\bmu_z(t), \Sigma_z(t)\bigr)$ is a Gaussian process characterized by
\begin{subequations}\label{eqn:ou-net-all}
\begin{align}
\bmu_z(t) &= e^{-(L_z+\eta I)t/\varepsilon}\bbE[\bZ(0)] \label{eqn:ou-net-mu} \\
\Sigma_z(t) &= e^{-(L_z+\eta I)t/\varepsilon}\bbE[\bZ(0)\bZ(0)^{\tr}]e^{-(L_z+\eta I)t/\varepsilon}
+ \gamma^2(L_z+\eta I)^{-1}\bigl(I - e^{-2(L_z+\eta I)t/\varepsilon}\bigr) . \label{eqn:ou-net-sigma}
\end{align}
\end{subequations}
Taking $t\to\infty$ in~\eqref{eqn:ou-net-all} yields the stationary distribution
$
\mu_{\infty}=\cN\bigl(\mathbf{0}, \gamma^2(L_z+\eta I)^{-1}\bigr). 
$
We can now consider~\eqref{eqn:w-dyn} defined with $\bZ_t$ governed by~\eqref{eqn:noise-network}, and average with respect to $\mu_{\infty}$:
\begin{align*}
dw_t &= -\bbE_{\mu_{\infty}}\Bigl\{\bigl(w_t\nor{\bx + \bZ_t}^2 - \scal{\bx + \bZ_t}{\by}\bigr)\Bigr\}dt + \sigma dB_t\\
&= -\Bigl[w_t\bigl(\nor{\bx}^2 + \gamma^2\trace(L_z+\eta I)^{-1}\bigr) -\scal{\bx}{\by}\Bigr]dt + \sigma dB_t
\end{align*}
where we have used that $\bbE[\nor{\bZ_t}^2] = \gamma^2\trace(L_z+\eta I)^{-1}$. As before, the averaged approximation is good when $\varepsilon\to 0$.
An expression identical to~\eqref{eqn:w-coupled},
\begin{equation}\label{eqn:sys-with-dep-noise}
d\bw_t = (L+\alpha I)(\bmu - \bw_t)dt + \sigma\dB_t
\end{equation}
 is obtained by redefining $\alpha:=\nor{\bx}^2 + \gamma^2\trace(L_z+\eta I)^{-1}$ and $\bmu:=(\scal{\bx}{\by}/\alpha)\bbone$. In this case,
 \[
 \lambda = \alpha - \nor{\bx}^2 = \gamma^2\trace(L_z+\eta I)^{-1} .
\]
Theorem~\ref{thm:sys-convergence} may be immediately applied to understand~\eqref{eqn:sys-with-dep-noise}. As before, the covariance of $\bZ_t$ figures into the regularization parameter. However now the covariance of $\bZ_t$ is a function of the network Laplacian $L_z=L_z(t)$, which is defined by the  topology and potentially {\em time-varying} coupling strengths of the noise network. 
By adjusting the coupling in~\eqref{eqn:noise-network}, we adjust the regularization $\lambda$ imposed upon~\eqref{eqn:sys-with-dep-noise}. 
When coupling increases, the dependence among the $Z_t^i$ increases and \mbox{$\trace(L_z+\eta I)^{-1}$} (and therefore $\alpha$) decreases. Thus, {\em increased correlation among observational noise variables implies decreased regularization}.

In the case of all-to-all coupling with uniform strength $\kappa\geq 0$, for example, $L_z$ has eigenvalues $0=\lambda_0<\lambda_1=\cdots=\lambda_m=m\kappa$. The regularization may in this case range over the interval 
\[
\inf_{\kappa} \trace(L_z+\eta I)^{-1} = \frac{1}{\eta} < 
\frac{\lambda}{\gamma^2}\leq
\frac{m}{\eta} = \sup_{\kappa} \trace(L_z+\eta I)^{-1}
\]
by adjusting the coupling strength $\kappa\in[0,\infty)$. Note that all-to-all coupling may be plausibly implemented with $\cO(n)$ connections using mechanisms such as {\em quorum sensing} (see~\citep{Taylor:Sci:09}, \citep[\S 2.3]{Bouvrie:NECO:11}).


\section{Regularization in predictive coding hierarchies}
\label{sec:learning-rao-ballard}

We now consider regularization and synchronization in the context of a two-layer, linear coding network. The particular system we will explore is representative of a broad category of computations occurring in multiple brain areas responsible for sensory information processing~\citep{Foldiak:90,OlshausenField:04,SmithLewicki:2006}. We will closely follow the setup proposed by~\cite{Rao:NNS:99}, where the first layer attempts to encode an input stimulus using a dictionary of basis functions and a top-down {\em prediction} of what the encoding should be, while the second layer attempts to encode the bottom layer's output using its own dictionary. The two layers view the input at different scales: the bottom layer takes high-resolution measurements using units with small receptive fields, while the top layer consists of units which pool over lower-layer units, giving effectively larger receptive fields at lower resolution at the top layer. The goal of our analysis will be to understand the role of noise and synchronization in and among such networks, and we will explore a network in which there is (1) noise affecting both top-down/bottom-up communication between layers, and (2) noise on the encoding dictionaries. The impact of, and interaction between, these noise sources is addressed, and we will show how communication noise may be reduced by synchronizing multiple hierarchies. Dictionary noise will be shown to have the effect of regularizing a suitable objective function, much like noise on the data provided an avenue for regularization in Sections~\ref{sec:noise-gives-reg} and~\ref{sec:calibrating-reg}.

\subsection{Model definition}
Rao and Ballard's model~\citep{Rao:NNS:99} assumes that a stimulus $\cI$ is represented in terms of a hierarchy of {\em causes} of increasing abstractness. At the bottom-most layer (which has access to the input), a coding variable $r$ is interpreted as a set of possible causes describing the input stimulus in terms of basis vectors stored in the columns of a dictionary $U$, so that $\cI=f(Ur) + \text{``noise''}$, where $f$ is a neural activation function. At the next, higher layer, the causes $r$ are in turn encoded in terms of a higher-level collection of basis elements described by the columns of $U_h$, so that $r=f(U_hz) + \text{``noise''}$, where $z$ is a vector representing a higher-level set of causes. The quantity $f(U_hz)$ is the {\em top-down prediction} of $r$.

Given {\em fixed} dictionaries $U,U_h$ at the bottom and top layers respectively, \cite{Rao:NNS:99} define  the problem of computing the encodings $r,z$ as that of jointly minimizing the (respective) bottom and top objective functions
\begin{subequations}\label{eqn:rb-objs}
\begin{align}
E_r(r,z) &:= \nor{\cI - f(Ur)}_2^2 + \lambda\nor{r-f(U_hz)}_2^2 \label{eqn:obj-botup} \\
E_z(r,z) &:= \nor{r - f(U_hz)}_2^2  \label{eqn:obj-topdn}
\end{align}
\end{subequations}
where $\lambda$ is a parameter controlling the trade-off between fidelity of the stimulus encoding and adherence to the top-down prediction. 
We will refer to the quantity $\bigl(\cI-f(Ur)\bigr)$ as the {\em bottom-up error signal}, and the quantity $\bigl(r-f(U_hz)\bigr)$ as the {\em top-down error signal}. From here on, we will make the simplifying assumption that $f$ is the identity map, corresponding to linear activation on the neural units. A linear activation $f$ produces pair of objectives $E_r,E_z$ which are quadratic in $r$ and $z$. In our analysis, the units at each layer are adapted to encode their inputs by way of gradient descent in continuous time, giving $\dot{r}=dr/dt = -\gamma\nabla_r E_r(r,z)$, $\dot{z}=dz/dt = -\gamma\nabla_z E_z(r,z)$. Setting $\gamma=1/2$ to eliminate the extraneous factor of 2 in the gradient gives the coupled pair of linear systems
\begin{subequations}
\label{eqn:rb-noisefree-system}
\begin{align}
\dot{r} &= U^{\tr}\cI - U^{\tr}Ur + \lambda(U^hz - r) \\
\dot{z} &= U_h^{\tr}r - U_h^{\tr}U_hz \,.
\end{align}
\end{subequations}
We will assume that the dictionaries $U,U_h$ remain fixed, and study the behavior of this ``coding system'' in the presence of noise.

\subsection{Noise on the error signals}
We now explore what happens when zero-mean Gaussian noise with covariance $\sigma^2 I,\sigma_h^2 I$ has been added to the bottom-up and top-down error signals appearing in~\eqref{eqn:obj-botup}-\eqref{eqn:obj-topdn}, respectively. Such noise might model both exogenous noise on the sensory input, as well as endogenous noise representing communication error, background neuronal noise, and/or quantization effects (due to, for instance, finite precision computation). The resulting continuous-time noisy gradient descent dynamics will be modeled as a system of It\^o stochastic differential equations. Taking the gradient of the objectives and allowing Brownian increments to play the role of the noise, the system~\eqref{eqn:rb-noisefree-system} can be augmented with appropriate diffusion terms and re-written as
\begin{equation}\label{eqn:rb-ito-sys}
d\begin{bmatrix}
r \\ z
\end{bmatrix} = 
\left(
\begin{bmatrix}
-(U^{\tr}U + \lambda I) & \lambda U_h \\
U_h^{\tr} & -U_h^{\tr}U_h
\end{bmatrix}
\begin{bmatrix}
r \\ z
\end{bmatrix} 
+ 
\begin{bmatrix}
U^{\tr}\cI \\ 0
\end{bmatrix}
\right)dt + 
\begin{bmatrix}
\sigma U^{\tr} & \lambda\sigma_h I \\
0 & \sigma_h U_h^{\tr}
\end{bmatrix} 
\begin{bmatrix}
dB_t \\ dB_t^h
\end{bmatrix} 
\end{equation}
where $B_t$ and $B_t^h$ are independent vector-valued Brownian motions.
In this form, we can see that the resulting system is solved by an OU process: a two-layer hierarchy with noise on the error signals can be expressed as a linear diffusion process. 

As discussed in Section~\ref{sec:gradient-learning}, synchronization among multiple coupled copies of such systems can reduce the variance of the noise appearing in~\eqref{eqn:rb-ito-sys}. Equivalently, synchronization can be used to reduce the variance or uncertainty surrounding the prediction/residual error signals in the Rao-Ballard model. In this sense, Needleman's notion of ``collective enhancement of precision'' \citep{Needleman:PD:2001} is a particularly apt interpretation, with synchronization playing a key role. 

We pursue this idea in more detail. Suppose that there are $M$ identical copies of the $d$-dimensional system~\eqref{eqn:rb-ito-sys}, each driven by independent Brownian motions, and coupled with a symmetric  matrix $W\in\bbR^{M\times M}$ of non-negative weights.  For ease of the notation, we will write each system in the coupled network using the generic notation
\begin{equation}
\label{eqn:rb-ito-sys-coupled}
dX_t^i = \left[c - AX_t^i + \sum_{j=1}^M W_{ij}(X_t^j - X_t^i)\right]dt + \Sigma dB_t^i \;, \qquad i=1,\ldots,M 
\end{equation}
where
\[
A:=
\begin{bmatrix}
U^{\tr}U + \lambda I & -\lambda U_h \\
-U_h^{\tr} & U_h^{\tr}U_h
\end{bmatrix}, \quad
c:=
\begin{bmatrix}
U^{\tr}\cI \\ 0
\end{bmatrix}, \quad
\Sigma:=\begin{bmatrix}
\sigma U^{\tr} & \lambda\sigma_h I \\
0 & \sigma_h U_h^{\tr}
\end{bmatrix}
\]
and $dB_t^i$ for each $i\in\{1,\ldots,M\}$ is an independent Brownian motion of the appropriate dimension.
Defining the graph Laplacian $L=\diag(W\bbone)-W$, we may write down a concise Langevin SDE governing the entire network of multiple diffusively coupled systems:
\begin{equation}
\label{eqn:rb-ito-sys-net}
dX_t = (L\oplus A)\bigl[(\bbone_M\otimes \mu) - X_t\bigr]dt + (I_M \otimes \Sigma)dB_t
\end{equation}
where ``$\otimes$'' denotes the matrix Kronecker product, $L\oplus A = L\otimes I_d + I_M\otimes A$ is the Kronecker direct sum of $L$ and $A$, and we have defined $X_t:=\bigl((X_t^1)^{\tr},\ldots,(X_t^M)^{\tr}\bigr)^{\tr}$, $B_t:=\bigl((B_t^1)^{\tr},\ldots,(B_t^M)^{\tr}\bigr)^{\tr}$, and $\mu=A^{-1}c$. The notation $I_d$ refers to the $d\times d$ identity matrix, and $\bbone_d$ refers to the $d$-dimensional vector of all ones. In this standard form, we can read off the equilibrium as $\bbone_M\otimes \mu$, and see immediately that the convergence of the noise-free dynamics is governed by $L\oplus A$.  The fact that $(L\oplus A)(\bbone_M\otimes \mu) = \bbone_M\otimes c$ may be verified by way of a simple calculation:
\begin{align*}
(L\oplus A)^{-1}(\bbone_M\otimes c) &= \left(\int_0^{\infty} e^{-(L\oplus A)t}dt\right)(\bbone_M\otimes c) \\
&= \int_0^{\infty} e^{-Lt}\bbone_M\otimes e^{-At}c \;dt \\
&= \bbone_M\otimes \int_0^{\infty}e^{-At}c \;dt \\
&= \bbone_M\otimes \mu \;.
\end{align*}
The first equality follows using the identity $\int_0^{\infty}e^{-Qt}dt=Q^{-1}$ for $Q$ positive definite, the second from applying the general property $e^{P\oplus Q}=e^P\otimes e^Q$ satisfied by the Kronecker sum, and the third from the fact that constant vectors are in the nullspace of the Laplacian.

The following result confirms that the variance of the noise affecting top-down/bottom-up signals in an individual system can be reduced by synchronizing multiple systems, each driven by independent noise.

\begin{theorem}
\label{thm:ou-sync-noisered}
Let $X_{\infty}:=\lim_{t\to\infty}X_t$ denote the stationary part of the solution to~\eqref{eqn:rb-ito-sys-net}, and let $\lmin$ denote the smallest non-zero eigenvalue of $L$. Then,
\[
\bbE\left[\frac{1}{M}\sum_{i=1}^M\nor{X_{\infty}^i-\mu}^2\right] \leq \frac{\lambda_{\text{max}}(\Sigma\Sigma^{\tr})}{2}
\left(\frac{d}{\lambda_{\text{min}}(A) + \lmin} + \frac{1}{M\lambda_{\text{min}}(A)}\right) .
\] 
\end{theorem}
This result parallels those found in~\citep{Tabareau10,Bouvrie:NECO:11}, and resembles Theorem~\ref{thm:sys-convergence} after initial transients. This is encouraging, as system~\eqref{eqn:rb-ito-sys-net} and the regression systems introduced above both take on Langevin form, and both arise from coupling multiple copies of a linear SDE. As in Section~\ref{sec:reg-sys-conv-behavior}, synchronization will cause individual trajectories to tend to the common equilibrium. We see that the first term in parentheses in Theorem~\ref{thm:ou-sync-noisered} can be made small through $\lmin$ by increasing the coupling strength and/or increasing the redundancy ($M$) depending on the network topology. The second term in parentheses comes from the fact that, for sufficiently strong coupling, the mean-field trajectory essentially has variance $\cO(1/M)$ around the equilibrium point. Clearly increasing $M$ reduces this variance.

\subsubsection{Stability}
\label{sec:rao-bal-stability}
In the continuous-time, stochastic setting we have considered, it is important to confirm that the systems of interest are in fact stable. We will show that system~\eqref{eqn:rb-ito-sys} is indeed stable in the sense of {\em stochastic contraction}~\citep{Pham09}. Stochastic contraction is a form of incremental stability for stochastic dynamical systems which seeks to bound the distance between any pair of trajectories of a system in expectation. We provide the required result from~\citep{Pham09} in a simplified form here for completeness.
\begin{theorem}{\citep[Theorem 2]{Pham09}}
\label{thm:stochastic-contraction}
Suppose $x_1(t),x_2(t)\in\bbR^d$ are two trajectories of the system
\[
dx=f(x,t)dt + \sigma(x,t)dW
\]
corresponding to the initial conditions $x(0)=x_1(0)$ and $x(0)=x_2(0)$, respectively, where $x_1(0),x_2(0)$ are chosen independently of the noise, and $f,\sigma$ satisfy the usual Lipschitz and growth conditions guaranteeing existence and uniqueness of solutions. If there exists a $\lambda > 0$ such that 
\begin{equation}
\label{eqn:contraction-condition}
\lambda_{\text{max}}\left(\frac{\partial f}{\partial x} + \frac{\partial f^{\tr}}{\partial x}\right) \leq -2\lambda
\end{equation}
and $\trace\bigl(\sigma(x,t)^{\tr}\sigma(x,t)\bigr)$ is uniformly upper-bounded by a constant $C$, then
\[
\bbE\bigl[\nor{x_1(t)-x_2(t)}^2\bigr] \leq \frac{C}{\lambda} + \bbE\bigl[\nor{x_1(0)-x_2(0)}^2\bigr]e^{-2\lambda t} \,,\quad \forall t \geq 0 .
\]
\end{theorem}
A system satisfying the condition~\eqref{eqn:contraction-condition} is said to be {\em contracting} with rate $\lambda$. The reader is referred to~\citep{Lohmiller98, WangSlotine05} for an introduction to contraction analysis and its connections to incremental stability, and to~\citep{Pham09} for details regarding the extension to stochastic systems. 

If multiple contracting systems are coupled sufficiently strongly through positive weights defining a strongly connected graph, then it is the case that the overall system is contracting (with rate determined by the smallest rate found among the individual systems); see~\citep{WangSlotine:06} and the references above. Hence, to show that~\eqref{eqn:rb-ito-sys-net} is stochastically contracting, it is enough to show that the noise-free version of~\eqref{eqn:rb-ito-sys} is contracting (that is, verify Equation~\eqref{eqn:contraction-condition}), and that the variance of the noise in~\eqref{eqn:rb-ito-sys} may be bounded uniformly in space and in time. There is no work involved in verifying the latter, since the diffusion coefficient depends on neither $(r,z)$ nor time. To verify that the noise-free part of~\eqref{eqn:rb-ito-sys} is contracting, we show that the symmetric part of the Jacobian is negative definite by equivalently showing that the (symmetrized) negated drift coefficient is positive definite.  This   latter quantity is given by
\begin{equation*}
\begin{bmatrix}
(U^{\tr}U + \lambda I) & -\frac{1}2{}(1+\lambda) U_h \\
-\frac{1}2{}(1+\lambda)U_h^{\tr} & U_h^{\tr}U_h
\end{bmatrix}.
\end{equation*}
This matrix is positive definite if and only if both $(U^{\tr}U + \lambda I)$ and the Schur complement 
\[
U_h^{\tr}U_h -  (\tfrac{1}{2}(1+\lambda))^2U_h^{\tr}(U^{\tr}U + \lambda I)^{-1}U_h 
\]
are both positive definite. The first is clearly positive definite (we assume $\lambda >0$). Rewrite the Schur complement as
$U_h^{\tr}(I-M)U_h$ with $M:=(\tfrac{1}{2}(1+\lambda))^2(U^{\tr}U + \lambda I)^{-1}$, and notice that the eigenvalues of $M$ are given by $\lambda_i(M)=(\tfrac{1}{2}(1+\lambda))^2/(s_i^2 + \lambda)$ where $s_i^2\geq 0$ is the $i$-th singular value of $U$. For the matrix $(I-M)$ to be positive definite,
we require $\lambda_i(M)\leq 1$. To avoid making specific assumptions about the spectrum of $U$, we may consider the sufficient condition $\lambda_i(M)\leq (\tfrac{1}{2}(1+\lambda))^2/\lambda\leq 1$, which has one solution at $\lambda= 1$.  Thus, if $\lambda=1$, $I-M$ is positive definite and there is a $C$ such that $C^{\tr}C=(I-M)$, in which case the Schur complement $U_h^{\tr}C^{\tr}CU_h$ is clearly recognized as positive definite too. We can therefore conclude that the SDE~\eqref{eqn:rb-ito-sys} is stochastically contracting by way of Theorem~\ref{thm:stochastic-contraction}, and this implies that the network~\eqref{eqn:rb-ito-sys-net} is also stochastically contracting.


\subsection{Noise on the dictionary}
\label{sec:rao-bal-dictnoise}
We now consider the setting in which, in addition to noise on the error signals, there is also additive noise affecting the fixed dictionaries used to encode signals in the hierarchy. We will show that this setting is closely related to the regression dynamics with noisy data explored above. It will be seen that noise on the dictionaries imposes regularization, and that synchronization plays an important role in calibrating this regularization.

We will proceed by replacing $U$ with the quantity $U+N_t$ and $U_h$ with $U_h+N_t^h$ in~\eqref{eqn:rb-ito-sys}, where $N_t, N_t^h$ are ergodic matrix-valued Gaussian noise processes respectively satisfying $\lim_{t\to\infty}\bbE[N_t]=\lim_{t\to\infty}\bbE[N_t^h]=0$ and $\lim_{t\to\infty}\bbE[N_t^{\tr}N_t] =: \Sigma_N, \lim_{t\to\infty}\bbE[(N_t^h)^{\tr}N_t^h] =: \Sigma_{N_h}$. As before, we will model the noise with a zero-mean OU noise process with correlation parameter $\varepsilon$ and consider an approximating dynamics valid for small $\varepsilon$ and timescales of $\cO(1)$. Applying the averaging result in Theorem~\ref{thm:averaging} to~\eqref{eqn:rb-ito-sys} modified as described, we have for the averaged diffusion coefficient
\begin{align*}
\bbE_{N,N^h}\left\{\begin{bmatrix}
\sigma U^{\tr} & \lambda\sigma_h I \\
0 & \sigma_h U_h^{\tr}
\end{bmatrix} 
\begin{bmatrix}
\sigma U^{\tr} & \lambda\sigma_h I \\
0 & \sigma_h U_h^{\tr}
\end{bmatrix}^{\tr} \right\} 
& =
\begin{bmatrix}
\sigma^2(U^{\tr}U + \Sigma_N) + \lambda^2\sigma_h^2 I &  \lambda\sigma_h^2 U_h \\
\lambda\sigma_h^2 U_h^{\tr} & \sigma_h^2(U_h^{\tr}U_h + \Sigma_{N_h})
\end{bmatrix}\\
 &= \Sigma\Sigma^{\tr} + \begin{bmatrix}
\sigma^2\Sigma_N & \\
 & \sigma_h^2\Sigma_{N_h}
\end{bmatrix} .
\end{align*}
Hence, the averaged system is given by
\begin{equation}\label{eqn:rb-avg-sys}
dX_t = -\left(A + 
\begin{bmatrix}
\Sigma_N & \\
 & \Sigma_{N_h}
\end{bmatrix}\right) X_t dt + cdt + \sqrt{\Sigma\Sigma^{\tr} + 
\begin{bmatrix}
\sigma^2\Sigma_N & \\
 & \sigma_h^2\Sigma_{N_h}
\end{bmatrix} 
 } dB_t ,
\end{equation}
where the square-root above is the matrix square-root, $X:=(r,z)^{\tr}$ and $A,c,\Sigma$ are as defined in~\eqref{eqn:rb-ito-sys-coupled} above. The system~\eqref{eqn:rb-avg-sys} is readily  seen as a noisy gradient dynamics minimizing the {\em regularized} objective 
\[
f(x) = x^{\tr}(A+D)x - x^{\tr}c \;,
\]
as its gradient with respect to $x$ is $\nabla_x f = (A+D)x - c$. Here, the matrix $D$ provides  the regularization. If $D$ is positive-definite, then it regularizes the problem of minimizing $g(x)=x^{\tr}Ax - x^{\tr}c$ by improving the conditioning of $A$. In the case of equation~\eqref{eqn:rb-avg-sys}, we see by inspection that the regularization is given by 
\begin{equation}
D = 
\begin{bmatrix}
\Sigma_N & \\
 & \Sigma_{N_h}
\end{bmatrix} .
\end{equation}
Clearly $D$ is positive definite since the noise covariances are. 
Moreover, the system~\eqref{eqn:rb-avg-sys} is easily shown to be stochastically contracting. The matrix $A$ in~\eqref{eqn:rb-avg-sys} is the (negated) diffusion coefficient appearing in~\eqref{eqn:rb-ito-sys}, and we can re-apply the reasoning in Section~\ref{sec:rao-bal-stability}. If the sufficient condition $\lambda:=1$ is satisfied, then $A$ is positive definite, and clearly $A+D$ will be positive definite too. Inspecting~\eqref{eqn:rb-avg-sys}, the diffusion coefficient is again independent of both time and the state. This fact, combined with positive definiteness of $A+D$, implies that~\eqref{eqn:rb-avg-sys} is stochastically contracting.

When the multidimensional OU-process governing the noise consists of independent components, $D$ will be diagonal. When these components are made dependent through synchronization between the components (but not between layers), $D$ will be block-diagonal, the coupling strength will influence the spectrum of $D$, and the regularization applied to $A$ may be controlled by adjusting the coupling. 

It is worth emphasizing that unlike the regression dynamics explored in Section~\ref{sec:noise-gives-reg}, here noise on the dictionary increases the variance of the noise on the gradient due to the second term under the square-root in Equation~\eqref{eqn:rb-avg-sys}. Indeed, noise on the dictionary can potentially amplify error signal noise due to a multiplicative, in addition to additive, interaction. A system with noisy dictionaries can be more sensitive to noise in top-down/bottom-up communications. If, however, multiple systems driven by independent noise processes are coupled, then the variance of the gradient noise can be reduced (approximately as $\cO(1/n)$) as explained in Section~\ref{sec:gradient-learning} and illustrated by the steady-state term in Equation~\eqref{eqn:full-error} of Theorem~\ref{thm:sys-convergence}.

Finally, we note that the analysis pursued in this section may be extended more generally to deeper hierarchies and other interconnections or assemblages of stochastic sub-systems of the type we have discussed. Stochastic contraction of the overall system follows from contraction of the individual systems~\citep{WangSlotine05}, however a global analysis of regularization properties is likely to be more difficult in the case of nonlinear dynamics or coupling.


\section{Distributed computation with noise}
\label{sec:distrib-computation}
We have argued that noise can serve as a mechanism for regularization. Noise may also be harnessed, in a different sense, to compute dynamics of the type discussed above. The distributed nature of the mechanism we will explore adheres to the general theme of parallel computation in the brain, and provides one possible explanation for how the gradients introduced previously might be estimated. The development below is closely related to that of the weight-perturbation technique in reinforcement learning~\citep{JabriFlower:92}, simultaneous perturbation stochastic gradient descent (SGD) ideas appearing in the stochastic approximation literature~\citep{Spall:TAC:92,KushnerBook}, and related applications in the adaptive optics~\citep{Vorontsov:97,Vorontsov:00} and robotics~\citep{BagnellSurvey:13} communities.

\subsection{Parallel stochastic gradient descent}
Let $J(\bu):\bbR^d\to\bbR$ be a locally Lipschitz Lyapunov cost functional we wish to minimize with respect to some set of  control signals $\bu(t)\in\bbR^d$. Gradient descent on $J$ can be described by the collection of flows
\[
\frac{du_i(t)}{dt}=-\gamma\frac{\partial J}{\partial u_i}(u_1,\ldots,u_d), \qquad i=1,\ldots,d .
\]
We consider the case where the gradients above are estimated via finite difference approximations of the form
\[
\frac{\partial J(\bu)}{\partial u_i} \approx \frac{J(u_1,\ldots,u_i+\delta u_i,\ldots,u_d)
- J(u_1,\ldots,u_i,\ldots,u_d)}{\delta u_i},
\]
where $\delta u_i$ is a small perturbation applied to the $i$-th input. {\em Parallel stochastic gradient descent} (PSGD) involves applying  i.i.d.~stochastic perturbations $\delta u_i$  simultaneously to all inputs in parallel, so that the gradients $\partial_i J(\bu)$ are estimated  as
\begin{equation}\label{eqn:psgd}
\frac{\partial J(\bu)}{\partial u_i} \approx \delta J\delta u_i, \qquad i=1,\ldots,d
\end{equation}
where
$
\delta J = J(u_1+\delta u_1,\ldots,u_i+\delta u_i,\ldots,u_d+\delta u_d)
- J(u_1,\ldots,u_i,\ldots,u_d) .
$
If $\delta u_i$ are symmetric random variables with mean zero and variance $\sigma^2$,
then $\sigma^{-2}\bbE[\delta J\delta u_i]$ is accurate to $\cO(\sigma^2)$~\citep{JabriFlower:92}.

\subsection{Analysis of a simple stochastic gradient model}
The parallel finite difference approximation~\eqref{eqn:psgd} suggests a more biologically plausible mechanism for implementing gradient dynamics. If the perturbations $\delta u_i$ are taken to be Gaussian i.i.d. random variables, we can model parallel stochastic gradient descent as an Ito process:
\begin{subequations}\label{eqn:fd-orig-sys}
\begin{align}
d\bu_t & = -\gamma\bigl[J(\bu_t + \bZ_t) - J(\bu_t)\bigr]\bZ_t dt,  &\quad \bu(0) &=u_0   \label{eqn:grad_proc}\\
d\bZ_t &= -\frac{1}{\varepsilon}\bZ_t dt + \frac{\sigma}{\sqrt{\varepsilon}}d\bB_t,  &\quad \bZ(0)&=z_0  \label{eqn:noise_proc}
\end{align}
\end{subequations}
where $\bB_t$ is a standard $d$-dimensional Brownian motion. Additive noise affecting the gradient has been omitted from~\eqref{eqn:grad_proc} for simplicity, and does not change the fundamental results discussed in this section. The perturbation noise $\bZ_t$ has again been modeled as a white-noise limit of Ornstein-Uhlenbeck processes~\eqref{eqn:noise_proc}. When $\varepsilon\to 0$,  Equation~\eqref{eqn:grad_proc} implements PSGD using the approximation given by Equation~\eqref{eqn:psgd} with $\delta u_i$ zero-mean i.i.d.~Gaussian random variables.

We will first proceed with an analysis of~\eqref{eqn:fd-orig-sys} in the particular case where $J$ is chosen from the quadratic family of cost functionals of the form
$J(\bu) = \bu^{\tr}A\bu$ where $A$ is a symmetric, bounded and strictly positive definite matrix\footnote{Without loss of generality we may assume $A$ is symmetric since the anti-symmetric part does not contribute to the quadratic form. In addition, objectives of the form $u^{\tr}Au + b^{\tr}u + c$ may be expressed in the homogeneous form $u^{\tr}Au$ by a suitable change of variables.}. In this setting the  analysis is simpler and suffices to illustrate the main points. This cost function satisfies  
$
\min_{\bu\in\bbR^d}J(\bu) = 0
$ with minimizer $\bu^{*}=0$, and $J$ is a Lyapunov function. Equation~\eqref{eqn:grad_proc} now takes the form
\begin{equation}\label{eqn:quad_dyn}
d\bu_t = -\gamma\bigl(2\bu_t^{\tr}A\bZ_t + \bZ_t^{\tr}A\bZ_t\bigr)\bZ_t dt, \qquad\qquad \bu(0)=u_0.
\end{equation}

What is the convergence behavior of~\eqref{eqn:quad_dyn}, and what is the precise role of the stochastic perturbations $\bZ_t$ used to estimate the gradients? These perturbations must be small in order to obtain accurate approximations of the gradients. However, one may also expect that the noise will play an important role in determining convergence properties since it is the noise that ultimately perturbs the system ``downhill'' towards equilibrium. 
Homogenizing~\eqref{eqn:quad_dyn} with respect to $\bZ_t$ leads to the following Theorem, the proof of which is given in the Appendix.
\begin{theorem}\label{thm:quad-pgsd-sys}
For any $0\leq t \leq T <\infty$, the solution $\bu(t)$ to~\eqref{eqn:quad_dyn} satisfies
\begin{equation}\label{eqn:quadratic_soln}
\lim_{\varepsilon\to 0}\bbE[\bu(t)] = e^{-\gamma\sigma^2At}\bu(0) .
\end{equation}
\end{theorem}
It is clear from this result that the PSGD system~\eqref{eqn:fd-orig-sys}, for  $\varepsilon\to 0$, converges in expectation  globally and exponentially to the minimum of $J$ when $J$ is a positive definite quadratic form. Our earlier intuition that the perturbation noise $\sigma$ should play a role in the rate of convergence is also confirmed: greater noise amplitudes lead to faster convergence. However this comes at a price. The covariance of $\bu(t)$ after transients is essentially the covariance of $\bZ_t$. Thus an inherent trade-off between speed and accuracy must be resolved by any organism implementing PSGD-like mechanisms. The analysis also suggests that a profitable learning strategy would be to start with large noise, and then anneal (reduce) the noise as the dynamics become close to the equilibrium point. In this case convergence will be accelerated, but the steady-state solution will be close to the true equilibrium point on average.  We will pursue the impact of PSGD noise on convergence and the steady-state variance in further detail below.

\subsection{Regression dynamics with the PSGD gradient}
We now turn to an analysis of the noisy, networked regression dynamics discussed in Section~\ref{sec:calibrating-reg}, reformulated to make use of the PSGD gradient approximation. Here, there is a strong case for synchronizing multiple systems, because while the gradient dynamics associated to a quadratic objective are linear, the PSGD dynamics are still quadratic in the variable of interest. In the  analysis of this system that follows, we will find that the regularization imposed upon the underlying optimization problem (and the means for adjusting it) is the same as in the gradient system, while the convergence rates and equilibrium variance are strongly influenced by the PSGD noise.

Let $\bigl(Q(\bw, \bx)\bigr)_i := E(w_i, \bx) = \tfrac{1}{2}\nor{\by - w_i\bx}^2$ for $i=1,\ldots,n$. As before, assume that the gradient dynamics are corrupted by ambient noise and the data is corrupted by its own (independent) additive noise. Letting $\bZ_t\in\bbR^m$ denote the data noise process, and $\bN_t\in\bbR^n$ denote the PSGD perturbation noise, the system we wish to study is given by
\begin{subequations}\label{eqn:coupled-system-spsa}
\begin{align}
d\bw_t &= -\gamma\bigl[Q(\bw_t + \bN_t, \bx + \bZ_t) - Q(\bw_t, \bx + \bZ_t)\bigr]\bN_tdt -L\bw_tdt + \sigma_w \dB_t^{(1)} \label{eqn:w-dyn-spsa}\\
d\bZ_t &= -\frac{1}{\varepsilon}(L_z + \eta I)\bZ_t dt + \frac{\sqrt{2}\sigma_Z}{\sqrt{\varepsilon}}\dB_t^{(2)},
\label{eqn:coupled-data-noise-spsa} \\
d\bN_t &= -\frac{1}{\varepsilon}\bN_t dt + \frac{\sqrt{2}\sigma_N}{\sqrt{\varepsilon}}\dB_t^{(3)}
\label{eqn:spsa-noise} 
\end{align}
\end{subequations}
where we have assumed that the Brownian motions above are all independent of each other. Substituting in $Q$ and simplifying reveals that each coordinate of $\bw_t$ is governed by dynamics of the form 
\begin{equation}\label{eqn:w-spsa-coord}
dw_t^i = -\frac{\gamma}{2}(N_t^i)^2(\bx + \bZ_t)^{\tr}\bigl[(\bx + \bZ_t)(2w_t^i+N_t^i)-2\by\bigr]dt - (L\bw_t)_idt + \sigma_w(\dB_t^{(1)})_i
\end{equation}
where $N_t^i$ denotes the $i$-th coordinate of $\bN_t$, and $w_t^i$ the $i$-th coordinate of $\bw_t$. To this expression we apply the averaging theorem, Theorem~\ref{thm:averaging}, twice: once averaging with respect to the stationary distribution of the data noise process, $\bZ_{\infty}\sim\cN\bigl(\mathbf{0}, \sigma_Z^2(L_z+\eta I)^{-1}\bigr)=:\mu_{\bZ_{\infty}}$, and again, averaging with respect to the stationary distribution of the PSGD noise $\bN_{\infty}\sim \cN\bigl(\mathbf{0}, \sigma_N^2I\bigr)=:\mu_{\bN_{\infty}}$. We have,
\begin{equation}
\begin{split}
\bbE_{\mu_{\bZ_{\infty}}}\bbE_{\mu_{\bN_{\infty}}}&\Bigl\{ (N_t^i)^2(\bx + \bZ_t)^{\tr}\bigl[(\bx + \bZ_t)(2w_t^i+N_t^i)-2\by\bigr] \Bigr\} \\
 &= \bbE_{\mu_{\bZ_{\infty}}}\Bigl\{
\nor{\bx + \bZ_t}^2(2w_t^i\sigma_N^2) - 2\sigma_N^2(\bx + \bZ_t)^{\tr}\by \Bigr\} \\
 &= 2\sigma_N^2\bigl[\bigl(\nor{\bx}^2 + \sigma_Z^2\trace(L_z+\eta I)^{-1}\bigr)w_t^i - \scal{\bx}{\by}\bigr].
\end{split}
\end{equation}
Applying the result of this calculation to Equation~\eqref{eqn:w-spsa-coord} and simplifying gives the averaged dynamics of $w^i$ as
\[
dw_t^i = \gamma\sigma_N^2\alpha(\mu - w_t^i)dt - (L\bw)_idt + \sigma_wdB_t^i
\]
where $\alpha:=\nor{\bx}^2 + \sigma_Z^2\trace(L_z+\eta I)^{-1}$ and $\mu:=\scal{\bx}{\by}/\alpha$. The dynamics of the network as a whole may be expressed in matrix-vector form as
\begin{equation}\label{eqn:w-final-spsa}
d\bw_t = (L + \gamma\sigma_N^2\alpha I)(\bmu - \bw_t)dt + \sigma_w\dB_t
\end{equation}
where $\bmu:=\mu\bbone$, making the similarity with Equation~\eqref{eqn:sys-with-dep-noise} clear. The regularization $\lambda = \alpha - \nor{\bx}^2 = \sigma_Z^2\trace(L_z+\eta I)^{-1}$ is the same as in Section~\ref{sec:calibrating-reg} and is not affected by the PSGD approximation, while the expected steady-state solution of Equation~\eqref{eqn:w-final-spsa} is also the same as that of Equation~\eqref{eqn:sys-with-dep-noise}. 

The use of PSGD gradients does, however, affect the rate of convergence towards equilibrium as well as the synchronization rate. We can apply Theorem~\ref{thm:sys-convergence} to Equation~\eqref{eqn:w-final-spsa} by setting $\nu=\gamma\sigma_N^2$ and choosing $\lambda$ as above, in which case we have that
\begin{equation}\label{eqn:spsa-sys-convergence}
\bbE\bigl[\tfrac{1}{n}\nor{\bw(t) - \bmu}^2\bigr] \leq
 \widetilde{C}e^{-2(\lmin + \gamma\sigma_N^2\alpha)t} +
 \overline{C}e^{-2\gamma\sigma_N^2\alpha t} +
 \frac{\sigma_w^2}{2}\left(\frac{1}{\lmin+\gamma\sigma_N^2\alpha} + \frac{1}{\gamma\sigma_N^2\alpha n}\right).
\end{equation}
Here, the synchronization rate is $2(\lmin + \gamma\sigma_N^2\alpha)$ and the rate of convergence of the mean-field trajectory is $2\gamma\sigma_N^2\alpha$. Hence, greater PSGD noise variance implies faster convergence. 

We also see from the third term in Equation~\eqref{eqn:spsa-sys-convergence} that the variance at equilibrium is apparently in inverse proportion to the variance of the PSGD noise: although the averaging procedure applied above reveals useful information about the behavior of the PSGD system on length scales of order $\cO(1)$ (and about the behavior of $\bw_t$ in expectation), it does {\em not} accurately capture detailed dynamics of this system on shorter length scales arising from the PSGD noise. This shortcoming can be explained by the fact that the PSGD noise enters into the dynamics of $\bw_t$ nonlinearly, through the objective function and again through a multiplication of the simultaneous finite-difference. In the particular case of~\eqref{eqn:coupled-system-spsa}, it will therefore be necessary to investigate the behavior of the solution process on a different length scale, in order to accurately characterize the impact of the PSGD noise on the second moment of $\bw_t$ at equilibrium. The following section explores this question in more detail.

\subsection{Impact of PSGD noise on the variance of the solution at equilibrium}
The previous section showed that the PSGD dynamics~\eqref{eqn:coupled-system-spsa} may be approximated by the coarse system~\eqref{eqn:w-final-spsa} on length scales of order $\cO(1)$. Reliance on~\eqref{eqn:w-final-spsa} as an approximation of~\eqref{eqn:coupled-system-spsa} leads to the (false) conclusion that the variance of the PSGD noise $\sigma_N^2$ can be used to improve convergence rates as well as the variance of the solution at equilibrium, and can therefore be made arbitrarily large. A similar paradox arises in a first order analysis of the weight perturbation algorithm (see e.g.~\cite[Sec. 17.3.1]{TedrakeNotes09}). In the present setting, we are led to this incorrect conclusion because terms of order $\cO\bigl((N_t^i)^3\bigr)$ in~\eqref{eqn:w-spsa-coord} have zero expectation. To reveal the contribution of such fast, zero-mean, higher order terms, we must study the effective dynamics on an appropriate small length scale. To this end, we will first average~\eqref{eqn:w-dyn-spsa} with respect to $\mu_{\bZ_{\infty}}$ and then rescale the resulting dynamics according to $\bw\to \bw\sqrt{\varepsilon}$. Finally, the resulting system will be homogenized, this time with respect to the PSGD noise, to obtain a simplified approximating SDE that can be analyzed towards providing insight into the original stochastic equations. This approach is useful when $\varepsilon$ is not taken to its zero-limit, but instead takes on more realistic (small but non-zero) values representing temporal correlations in the noise sources.

Averaging~\eqref{eqn:w-dyn-spsa} with respect to the data noise gives
\[
dw_t^i = -\frac{\gamma}{2}\bigl[2\alpha w_t^i(N_t^i)^2 + \alpha(N_t^i)^3 -2(N_t^i)^2\scal{\bx}{\by}\bigr]dt
-(L\bw_t)_idt + \sigma_w(\dB_t^{(1)})_i,\quad i=1,\ldots,n
\]
where we define $\alpha:=\nor{\bx}^2 + \sigma_Z^2\trace(L_z+\eta I)^{-1}$ and $\mu:=\scal{\bx}{\by}/\alpha$ as before.
Since this system is autonomous, we can simplify the calculations by shifting the averaged dynamics to have a  steady state expectation of zero. Combining this shift with the rescaling above leads to the change of variable $\bw\to\sqrt{\varepsilon}\bw + \mu\bbone$, giving
\begin{equation}\label{eqn:rescaled-spsa-w}
dw_t^i = -\frac{\gamma}{2}\bigl[2\alpha w_t^i(N_t^i)^2 + \varepsilon^{-1/2}\alpha(N_t^i)^3\bigr]dt
-(L\bw_t)_idt + \varepsilon^{-1/2}\sigma_w(\dB_t^{(1)})_i,\quad i=1,\ldots,n
\end{equation}
recalling that $L\bbone=0$. 

We will again apply the homogenization results described in~\citep{PardVeretI} to obtain a simplified SDE approximating the behavior of~\eqref{eqn:rescaled-spsa-w}. In contrast to Section~\ref{sec:averaging-regression-system} however, the SDE to be averaged, Equation~\eqref{eqn:rescaled-spsa-w}, includes both drift and diffusion terms of order $\cO(\varepsilon^{-1/2})$. To handle this case, we will invoke Theorem 3 in~\citep{PardVeretI} rather than the simplified result described by Theorem~\ref{thm:averaging} above. The reader is referred to~\citep{PavStuBook} and references therein for a general discussion of homogenization theory for SDEs in which the equations to be averaged have terms of order $\cO(\varepsilon^{-1/2})$. Before proceeding, a final complication needs resolving. Homogenization theorems of the type presented in~\cite{PardVeretI} are typically stated in a form which requires the diffusion coefficient to be $\cO(1)$. Here, the noise appearing in~\eqref{eqn:rescaled-spsa-w} is $\cO(\varepsilon^{-1/2})$, so to get around this difficulty we will approximate the diffusion term by a fast colored noise process as we have done elsewhere. In this form, the homogenization theory cited above can be applied without modification. With this approximation, the resulting multiscale system may be expressed as
\begin{subequations}
\begin{align}
dw_t^i &= -\bigl[\gamma\alpha w_t^i(N_t^i)^2 + (L\bw_t)_i\bigr]dt + 
\frac{1}{\sqrt{\varepsilon}}\left[-\frac{\gamma\alpha}{2}(N_t^i)^3 + U_t^i\right]dt,\quad i=1,\ldots,n 
\label{eqn:spsa-rescaled-sys} \\
d\bN_t &= -\frac{1}{\varepsilon}\bN_t dt + \sqrt{\frac{2}{\varepsilon}}\sigma_N \dB_t^{(3)}\\
d\bU_t &= -\frac{1}{\varepsilon}\bU_t dt + \sqrt{\frac{2}{\varepsilon}}\sigma_w \dB_t^{(4)} \;.
\end{align}
\end{subequations}
For simplicity, we have assumed that a single correlation parameter $\varepsilon$ describes the two noise sources. Application of~\citep[Theorem 3]{PardVeretI} involves, in addition to computing expectations, setting up and solving a Poisson PDE. We provide only the final results here in the form of a Proposition, the details of which are deferred until the Appendix.

\begin{proposition}
\label{prop:spsa-homogenization}
For $\varepsilon\ll 1$ and times $t$ up to $\cO(1)$ the solution to~\eqref{eqn:spsa-rescaled-sys} is approximated by the solution $\bW_t$ to
\[
d\bW_t = -(L + \gamma\sigma_N^2\alpha I)\bW_t + \sqrt{2\sigma_w^2 + \tfrac{11}{2}\gamma^2\alpha^2\sigma_N^6}\dB_t \,.
\]
Furthermore, an estimate of the stationary variance of~\eqref{eqn:w-dyn-spsa} on length scales of order $\cO(\sqrt{\varepsilon})$ may be obtained by applying Theorem~\ref{thm:ou-sync-noisered} to this approximating SDE: For $0<\varepsilon\ll 1$ and times $t$ up to $\cO(1)$, we have
\[
\bbE\Bigl[\tfrac{1}{n}\nor{\bw_{\infty}-\mu\bbone}^2\Bigr] \leq \varepsilon\bigl(\sigma_w^2 + \tfrac{11}{4}\gamma^2\alpha^2\sigma_N^6\bigr)
\left(\frac{1}{\lmin + \gamma\sigma_N^2\alpha} + \frac{1}{\gamma\sigma_N^2\alpha n}\right) \,.
\] 
\end{proposition}

Whereas Theorem~\ref{thm:sys-convergence} applied to~\eqref{eqn:w-final-spsa} did not reveal the impact of PSGD noise on short length scales, we see that Proposition~\ref{prop:spsa-homogenization} above does provide this information by way of the additional term $\tfrac{11}{4}\gamma^2\alpha^2\sigma_N^6$ added to $\sigma_w^2$ on the right-hand side of the bound above. From this bound, the PSGD noise variance is seen to enter according to $\cO(\sigma_N^6/\sigma_N^2)=\cO(\sigma_N^4)$, confirming that one cannot make the perturbations arbitrarily large without increasing the variance at equilibrium.

\section{Simulations}
\label{sec:simulations}
In this Section we present simulations demonstrating the regularization ideas discussed above. We first describe experiments showing regularization (and the calibration thereof) in the context of the simple regression problem discussed in Sections~\ref{sec:gradient-learning}-\ref{sec:calibrating-reg}. Parallels with physiological data are highlighted. We then consider experiments involving predictive coding hierarchies of the type described in Section~\ref{sec:learning-rao-ballard}, in which the effects of feedback communication noise and dictionary noise are explored, and synchronization of multiple copies is confirmed to reduce communication uncertainty. Finally, we end the Section with an illustrative example of the PSGD gradient approximation in the context of a network of coupled, {\em nonlinear} stochastic systems performing gradient descent on a (non-convex) double-well objective, where analytical study would be more difficult. In particular, we consider the interaction between PSGD noise and gradient noise, and the impact this has on which of multiple local minima is selected.

\subsection{Regularized regression dynamics with connections to visual motion perception}
\begin{figure}[t]
\centering
\includegraphics[width=0.49\textwidth]{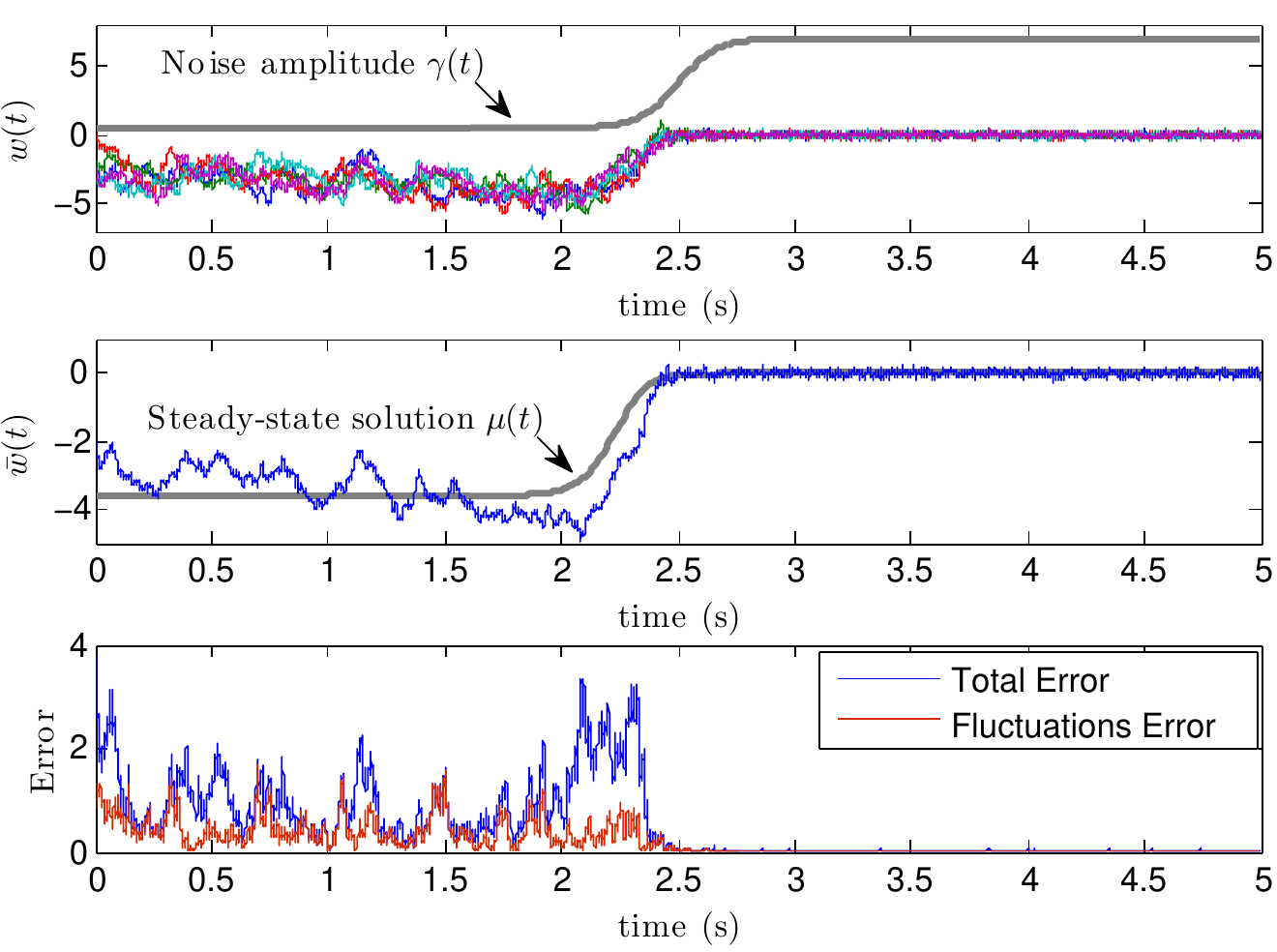}
\hskip 0.05cm
\includegraphics[width=0.49\textwidth]{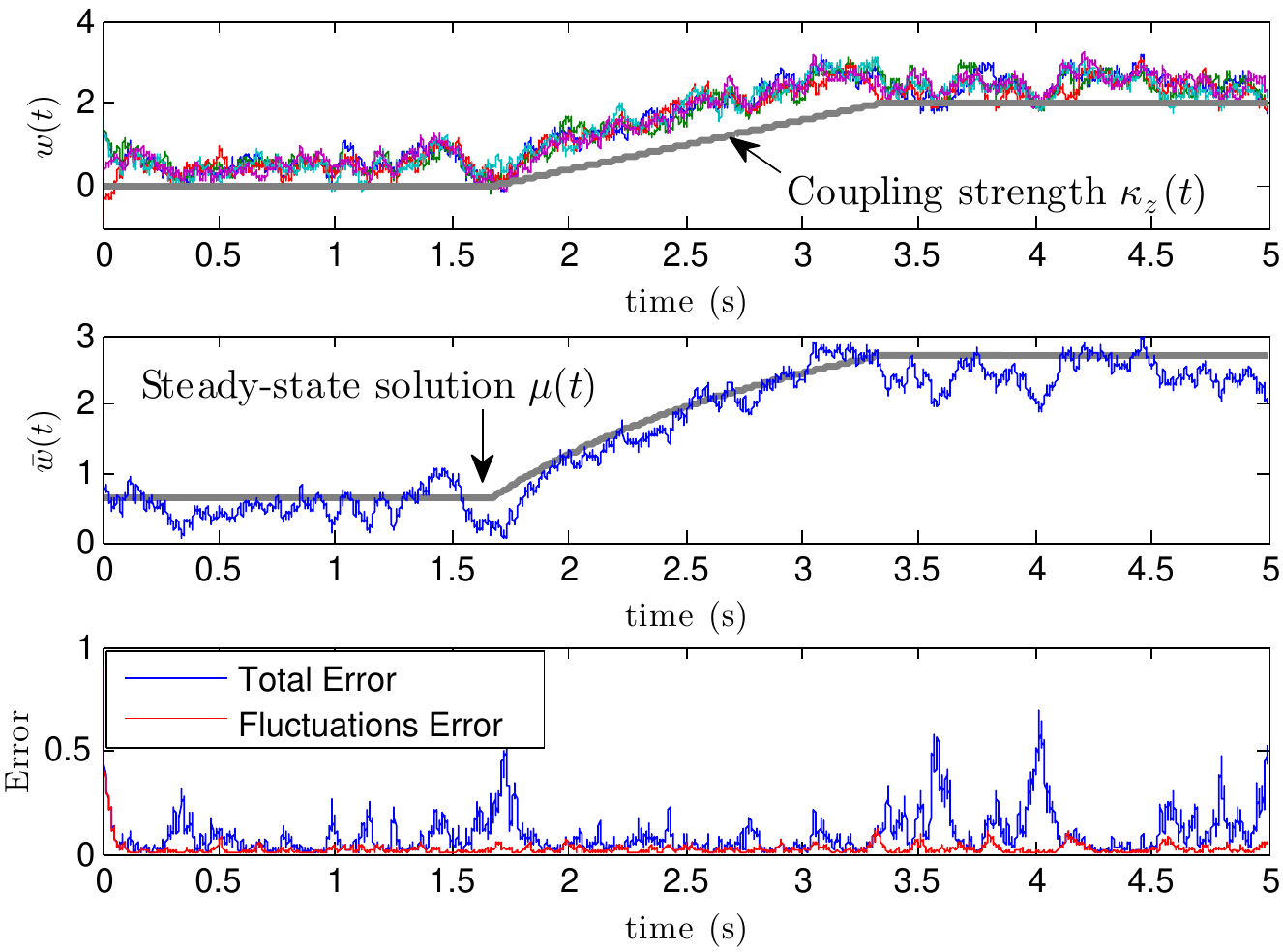}
\caption{{\small (Left stack) Increased observation noise imposes greater regularization, and leads to a reduction in ambient noise. (Right stack) Stronger coupling/correlation between observation noise processes decreases regularization. See text for details.}}
\label{fig:sims}
\end{figure}

We first simulated a network of gradient dynamics with uncoupled observation noise processes obeying~\eqref{eqn:uncoupled-noise-system}. To illustrate the effect of increasing observation noise variance, the parameter $\gamma$ in~\eqref{eqn:ou-noise} was increased from 0.5 to 7 along a monotonic, sigmoidal path over the duration of the simulation.  We used $n=5$ systems~\eqref{eqn:w-dyn} with $\sigma=4$, coupled all-to-all with uniform strength $\kappa=2$. Observations were sampled according to $(\bx)_i\sim\cN(0, 0.04)$, $(\by)_i\sim\text{Uniform}[0,20]$ with $m=20$ entries, once and for all, at the beginning of the experiment. Initial conditions were drawn according to $\bw(0)\sim\text{Uniform}[-3,3]$, and $\bZ(0)$ was set to $0$. Figure~\ref{fig:sims} (left three plots) verifies some of main conclusions of Section~\ref{sec:reg-remarks}. The top plot shows the sample paths $\bw(t)$ and time course of the observational noise deviation $\gamma(t)$ (grey labeled trace). When the noise increases near $t=2.5s$, a dramatic drop in the variance of $\bw(t)$ is visible. The middle plot shows the center of mass (mean-field) trajectory $\bar{w}(t)$ superimposed upon the time-varying noise-free solution $\mu(t)$ (gray labeled trace). Because the observation noise is increasing, the regularization $\lambda=m\gamma^2$ increases and the solution $\mu(t)$ to the regularized problem decreases in magnitude following~\eqref{eqn:reg_soln}. The bottom plot shows the mean-squared distance to the time-dependent noise-free solution $\bmu(t)$, and the mean-squared size of the fluctuations about the center of mass $\bar{w}$\footnote{These quantities are similar to those defined in~\eqref{eqn:sys-err-decomp}, but represent only this single simulation -- not in expectation. Here, ergodic theory allows one to (very roughly) infer ensemble averages by visually estimating time averages.}. It is clear that the error rapidly drops off when $\gamma(t)$ increases, confirming the apparent reduction in the variance of $\bw(t)$ in the top plot.

A second experiment, described by the right-hand stack of plots in Figure~\ref{fig:sims}, shows how synchronization can function to adjust regularization over time. This simulation is inspired by the experimental study of noise correlations in cortical area MT due to~\cite{Lisberger:09}, where it was suggested that time-varying correlations between pairs of neurons play a significant role in explaining behavioral variation in smooth-pursuit eye movements. In particular, the findings in~\citep{Lisberger:09} and~\citep{deOliveira:97} suggest that short-term increases in noise correlations are likely to occur after feedback arrives and neurons within and upstream from MT synchronize. We simulated a collection of correlated observation noise processes obeying~\eqref{eqn:noise-network} ($\varepsilon=10^{-3}$, $\eta=3$) with all-to-all topology and uniform coupling strength $\kappa_z(t)$ increasing from 0 to 2 along the profile shown in Figure~\ref{fig:sims} (top-right plot, labeled gray trace). This noise process $\bZ_t$ was then fed to a population of $n=5$ units obeying~\eqref{eqn:w-dyn}, with ambient noise $\sigma=1$ and all-to-all coupling at fixed strength $W_{ij}=\kappa=2$. New data $\bx,\by$ and initial conditions were chosen as in the previous experiment. The middle plot on 
the right-hand side shows the effect of increasing synchronization among the observation noise processes. As the coupling increases, the noise becomes more correlated and regularization decreases. This in turn causes the desired solution $\mu(t)$ to the regression problem to increase in magnitude (labeled gray trace). With decreased regularization, the ambient noise is more pronounced. The bottom-right plot shows the mean fluctuation size and distance to the noise-free solution (total error). An increase in the noise variance is apparent following the increase in observational noise correlation.
 

\begin{figure}[t]
\centering
\includegraphics[width=0.8\textwidth]{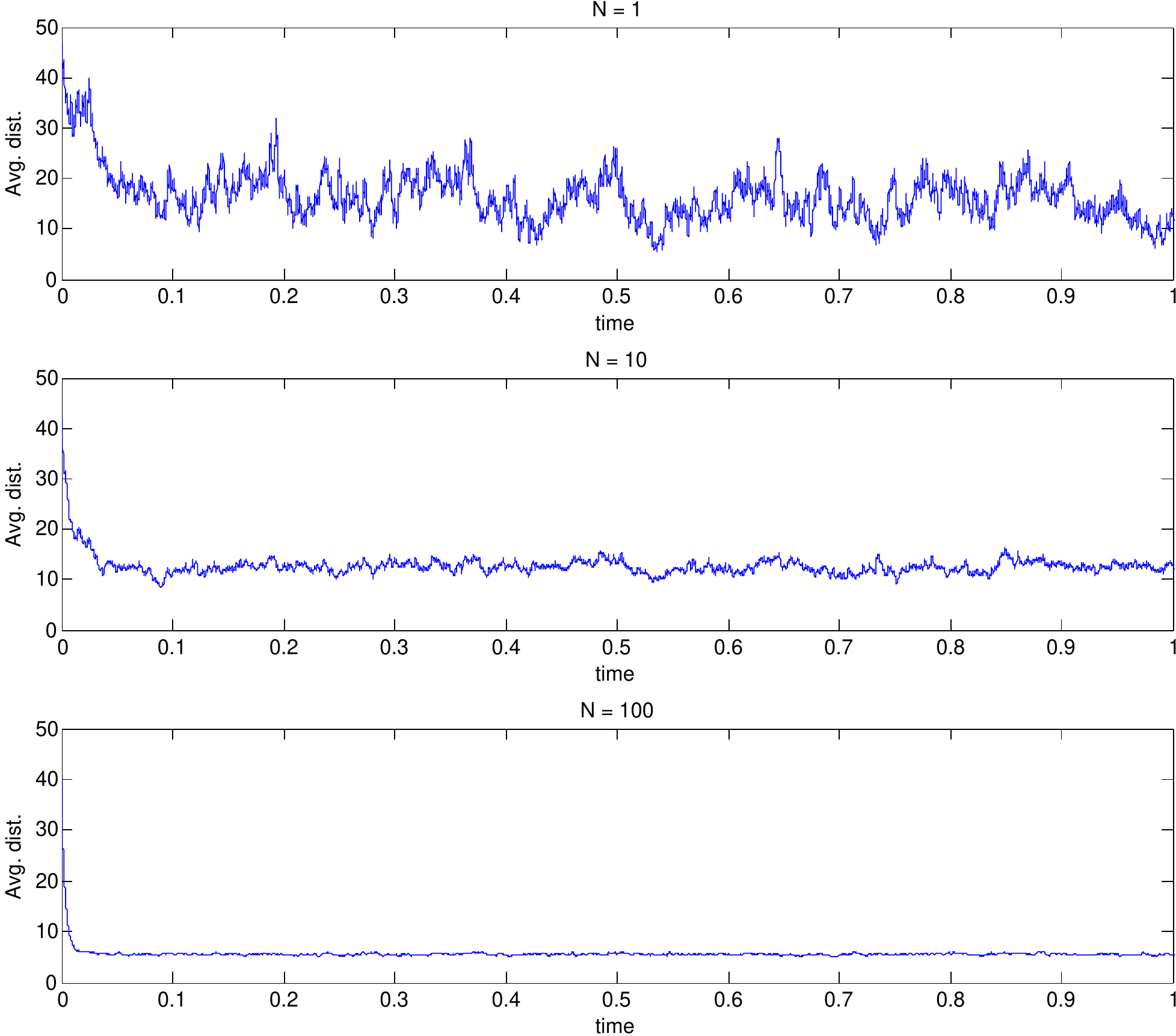}
\caption{\small Simulation of synchronized two-layer linear Rao-Ballard networks with noisy top-down/bottom-up error signals. The plots show the average distance between trajectories and the expected equilibrium over time for $N=1,10,100$ coupled copies of the network, assuming all-to-all coupling with strength $\kappa=1$.}
\label{fig:rb-noise-err-signal}
\end{figure}

\subsection{Hierarchical coding with noisy error signals}
\label{sec:expt-rb-error-signal} 

To illustrate how synchronization can reduce the effect of noise on the top-down and bottom-up error signals in a coding hierarchy, we simulated the network of coupled, two-layer Rao-Ballard hierarchies described by Equation~\eqref{eqn:rb-ito-sys-net} with $N=1,10,100$ copies. All-to-all (diffusive) coupling with strength $\kappa=1$ was used, giving  a minimum non-trivial Laplacian eigenvalue of $\lmin=N\kappa$. The following parameters were chosen to define the Rao-Ballard hierarchy: input dimension $\dim(\cI)=100$, bottom layer encoding dimension $\dim(r)=20$, top layer coding dimension $\dim(z)=10$, bottom layer noise deviation $\sigma=1$, top layer noise deviation $\sigma_h=1$, error feedback trade-off parameter $\lambda=1$.  Euler-Maruyama integration with step size $\Delta t=10^{-4}$ was used to integrate the SDE from $t=0$ to $t=1$. Trajectories were initialized with randomly drawn values from the uniform distribution on $[-2,2]$. Both the top and bottom layer encoding dictionaries were initialized randomly with elements drawn i.i.d. from the standard Normal distribution, but were kept fixed across experiments.

Figure~\ref{fig:rb-noise-err-signal} shows the average distance to equilibrium, $\tfrac{1}{N}\sum_i\nor{X_t^i-\mu}^2$, for simulations in which there were $N=1,10$ or $100$ copies (in top to bottom order, respectively).  It can be seen from the top plot that even error-feedback noise of modest variance $\sigma=\sigma_h=1$ imposes substantial noise on the system's trajectories due to amplification by the dictionaries. The precise relationship is explained by the diffusion coefficient of Equation~\eqref{eqn:rb-ito-sys}. Bearing in mind that we generated random dictionaries with normally distributed entries, from asymptotic random matrix theory the largest singular value of $U\in\bbR^{100\times 20}$ is approximately $\sqrt{100}+\sqrt{20}\simeq 14.47$ \citep{Rudelson:10}. Hence, amplification of the ambient noise imposed by the dictionaries may not be insignificant. If multiple copies are coupled so that they synchronize, then the impact of the noise is reduced. The middle and bottom plots confirm that this is indeed the case, where it can be seen that the variance is reduced and the convergence rate to equilibrium is also improved. 

\begin{figure}[t]
\centering
\includegraphics[width=0.48\textwidth]{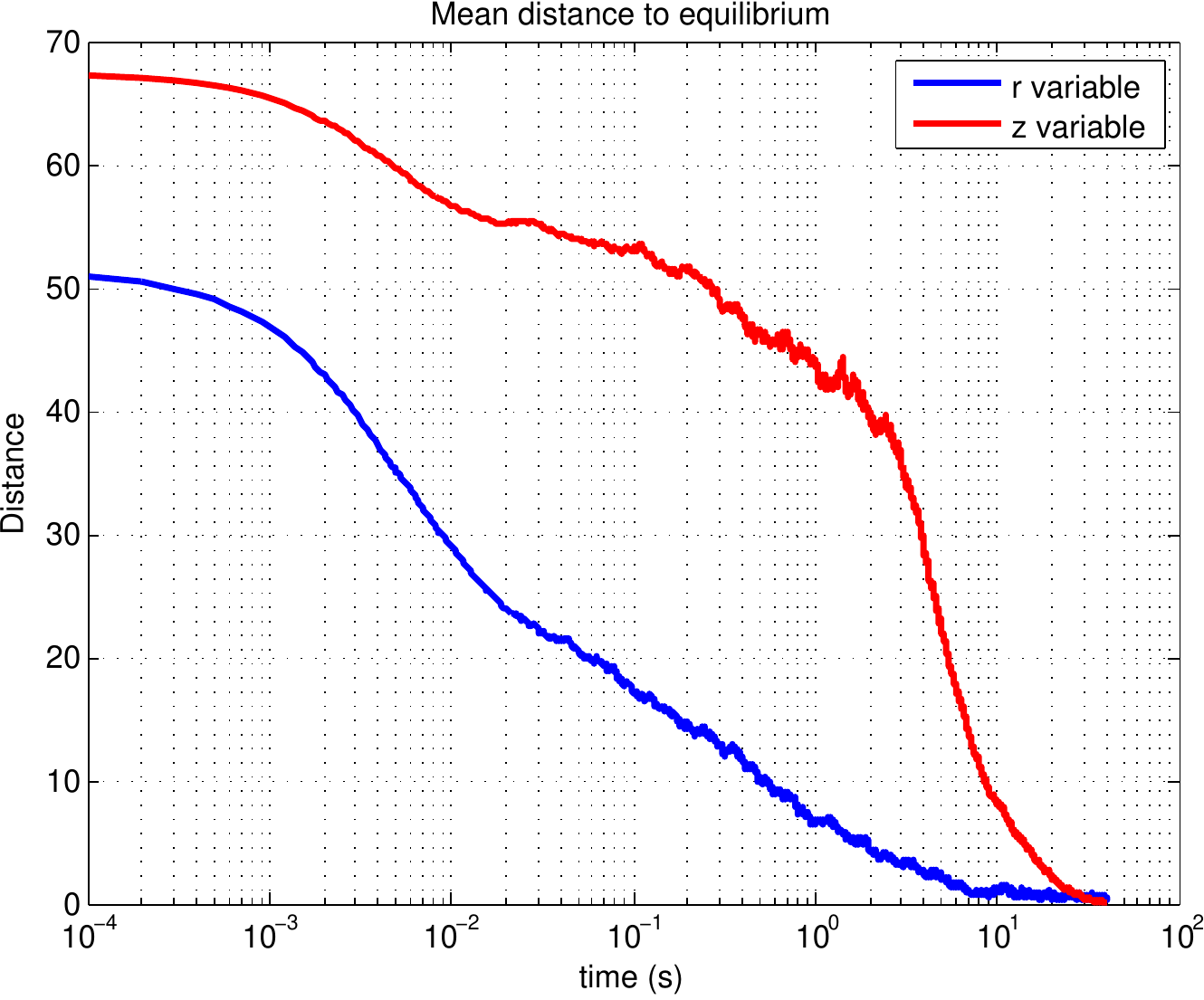}
\hspace{0.05cm}
\includegraphics[width=0.49\textwidth]{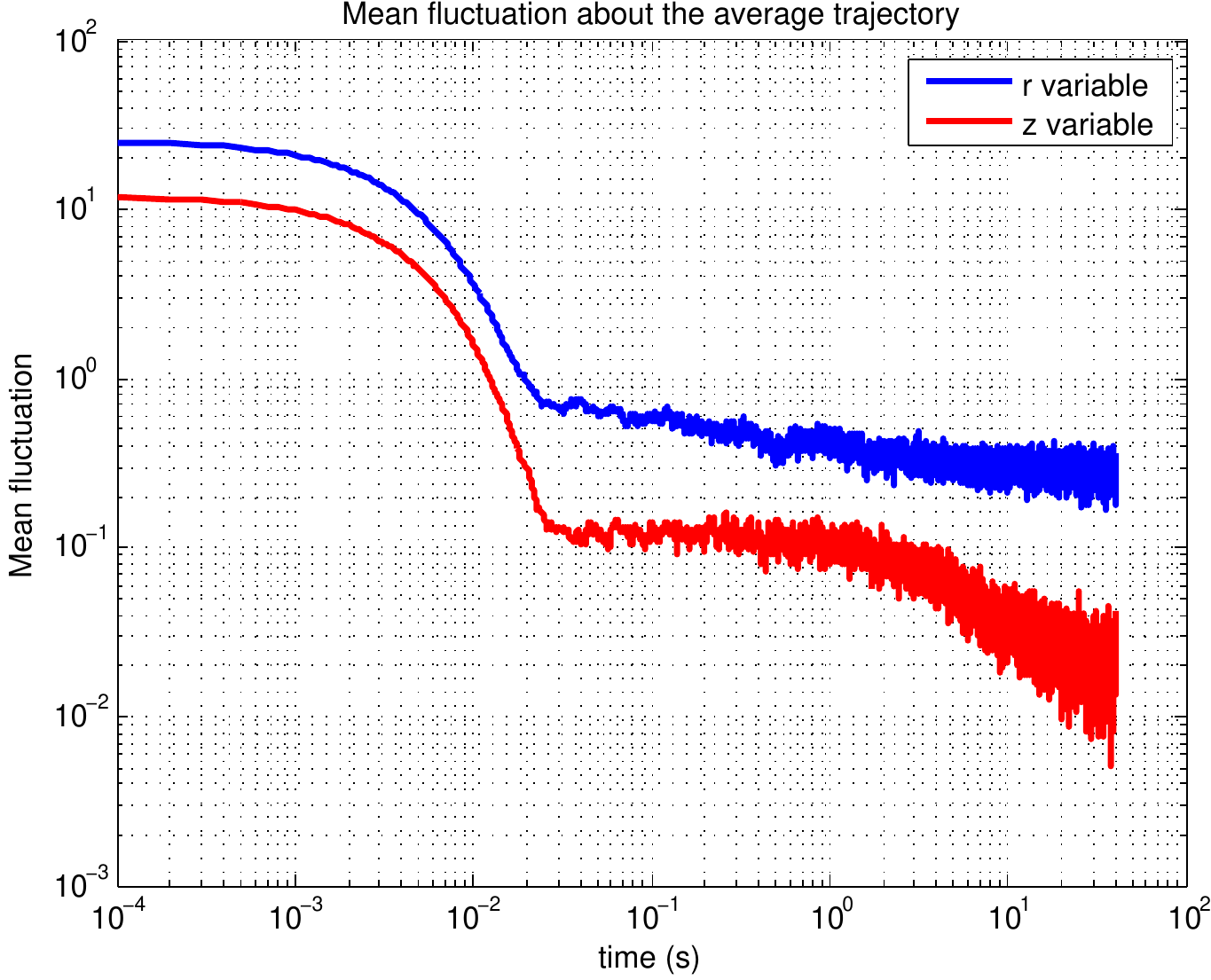} \\
\vspace{0.2cm}
\includegraphics[width=0.98\textwidth]{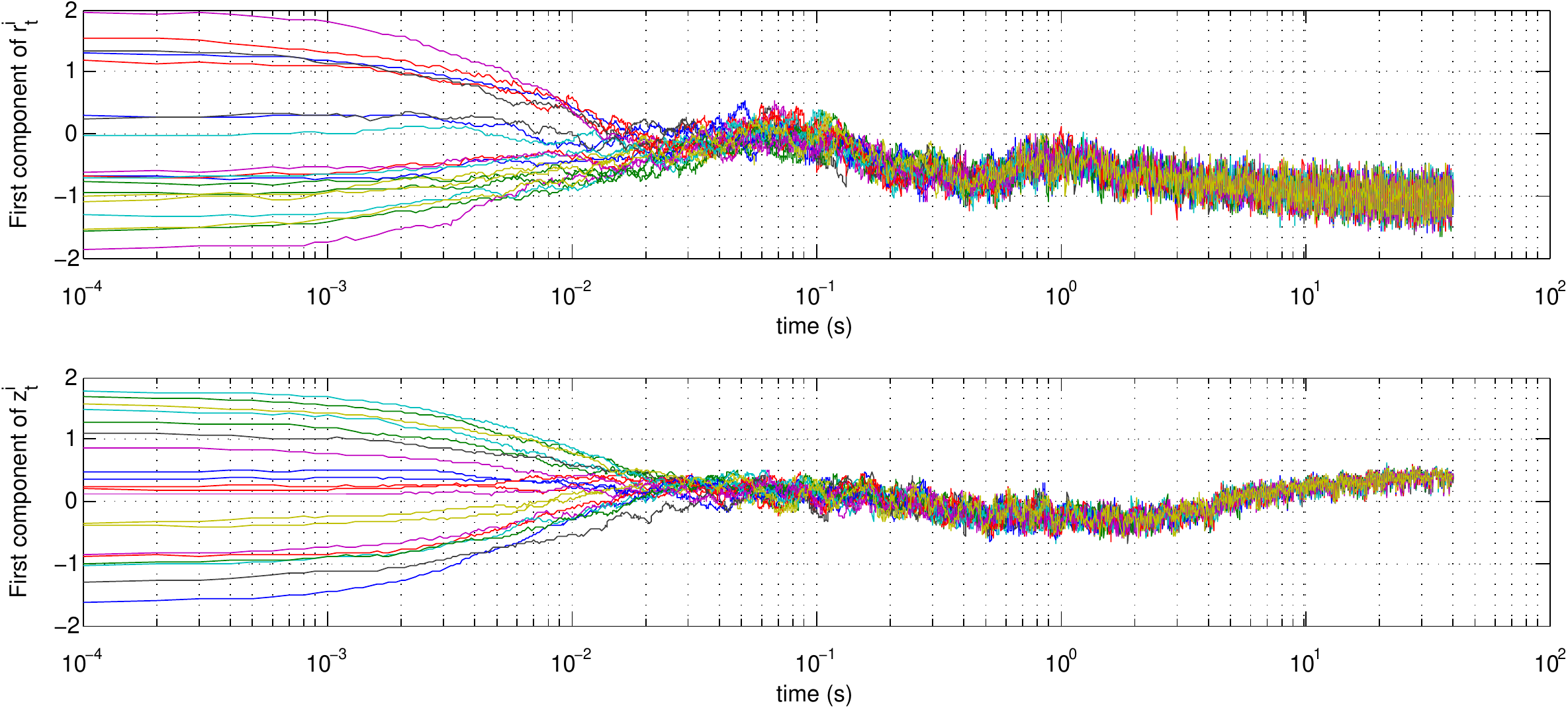}
\caption{\small Simulation of synchronized two layer nonlinear Rao-Ballard networks with noisy top-down/bottom-up error signals. Activation functions are the $\tanh$ nonlinearity, $N=20$, and coupling is all-to-all with strength $\kappa=5$.}
\label{fig:rb-nonlin}
\end{figure}

Lastly, we consider a collection of coupled {\em nonlinear} Rao-Ballard systems with noisy feedback signals, where reduction of noise variance by averaging requires synchronization. Here, we simulated the gradient dynamics implied by adding noise to the error signals in Equation~\eqref{eqn:rb-objs} and taking the gradient as before, but this time choosing $f(x) = \tanh(x)$ -- a sigmoidal activation nonlinearity. The resulting network of coupled systems is given by
\begin{align*}
dr_t^i &= F(r_t^i,z_t^i)dt + \kappa\sum_{j=1}^N(r_t^j-r_t^i)
+ \alpha(r_t^i)
\begin{bmatrix}
dB_t^i \\ dB_t^{h,i}
\end{bmatrix} \\
dz_t^i &= G(r_t^i,z_t^i)dt + \kappa\sum_{j=1}^N(z_t^j-z_t^i) + \beta(z_t^i)dB_t^{h,i}, \qquad i=1,\ldots,N
\end{align*}
where
\begin{align*}
F(r,z) &:= U^{\tr}\diag\{\bbone-f^2(Ur)\}\bigl(\cI - f(Ur)\bigr) + \lambda\bigl(f(U_hz) -r\bigr) \\
\alpha(r) &:= \bigl[U^{\tr}\diag\{\bbone-f^2(Ur)\}\sigma ~~~~ \lambda\sigma_h I\bigr] \\
G(r,z) &:= U_h^{\tr}\diag\{\bbone-f^2(U_hz)\}\bigl(r - f(U_hz\bigr)) \\
\beta(z) &:= U_h^{\tr}\diag\{\bbone-f^2(U_hz)\}\sigma_h 
\end{align*}
and the function $f$ is assumed to apply element-wise when passed a vector argument. 
For these experiments we took $N=20$, set $\sigma=\sigma_h=0.5, \kappa=5$ and simulated the system over $0\leq t \leq 40s$. All other parameters and experimental settings were taken as in the previous simulation above. Figure~\ref{fig:rb-nonlin} shows the results. The top left plot gives the squared Euclidean distance to equilibrium averaged over the copies  at each point in time, with separate traces for the $r$ and $z$ variables. The top right plot shows the squared norm of the fluctuations in $r$ and $z$ about the (respective) mean-field trajectories averaged over the copies at each point in time,
\[
\text{flucts}_r(t) = \frac{1}{N}\sum_{i=1}^N\Bigl\|r_t^i - \frac{1}{N}\sum_j r_t^j\Bigr\|^2 
\]
with a similar definition for the fluctuations in $z_t$. The plot shows that the systems approach the synchronization subspace with a comparatively small time constant. The bottom plots illustrate what the trajectories look like, and give the first component of each copy's $r^i_t$ variable (first plot) and $z^i_t$ variable (bottom plot) over time. Note that in these plots we have used logarithmic axes where appropriate to show detail. These traces give an alternate view of the synchronization early on, and then the variance later into the simulation after transients.

\begin{figure}[tp]
\centering
\includegraphics[width=0.9\textwidth]{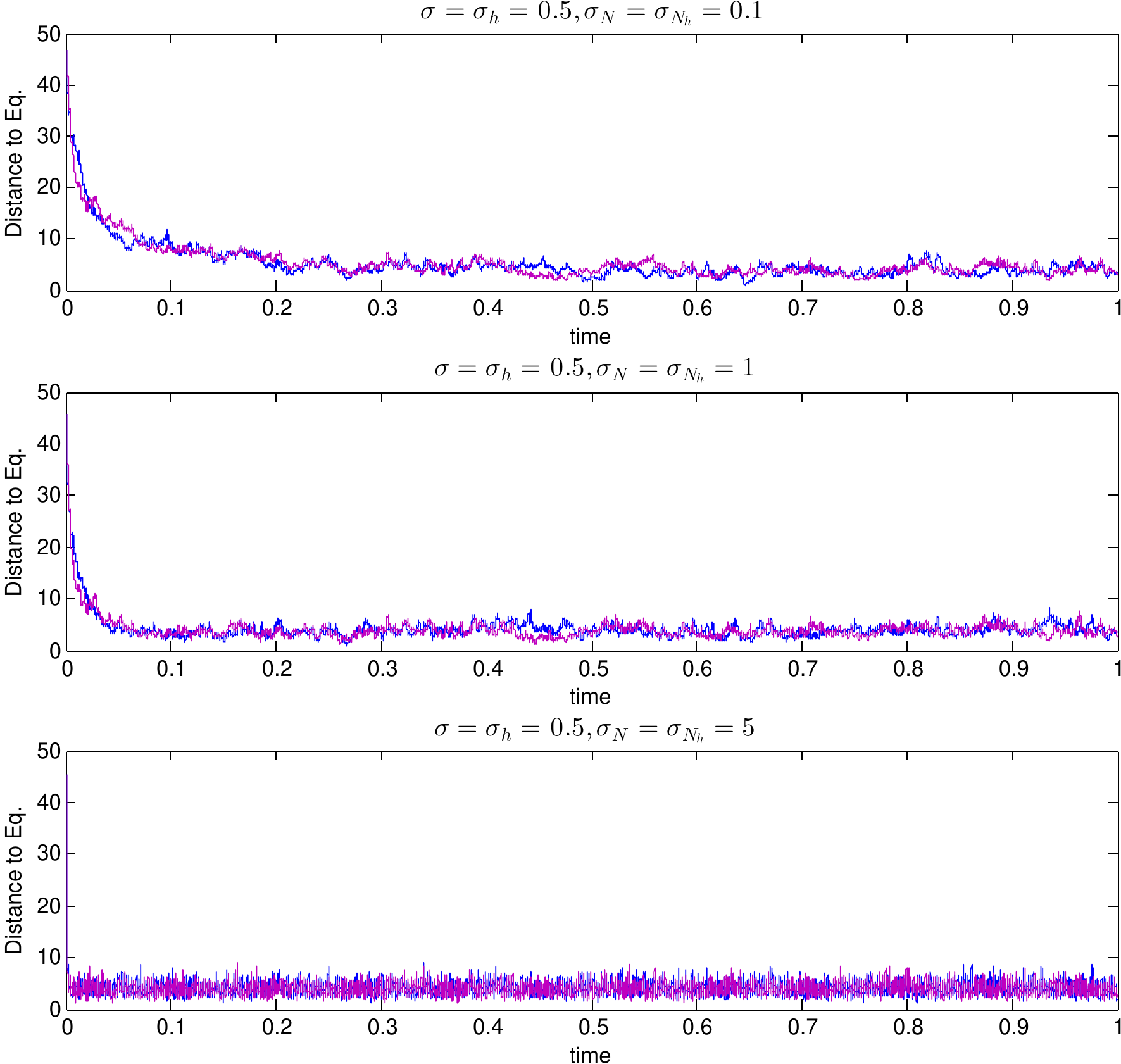}
\caption{\small Simulation of two layer Rao-Ballard networks with noisy top-down/bottom-up error signals and noisy dictionaries. The plots show distance from the mean-field trajectory to the expected equilibrium over time for different dictionary noise variances. \textcolor{violet}{\bf Purple} traces indicate trajectories of the averaged system~\eqref{eqn:rb-avg-sys} while \textcolor{blue}{\bf blue} traces correspond to the original system.}
\label{fig:rb-dictnoise}
\end{figure}

\subsection{Hierarchical coding with noisy dictionaries}
Following the development in Section~\ref{sec:rao-bal-dictnoise}, we considered the impact of additional noise on the coding dictionaries and simulated all-to-all networks of coupled systems of the form given by Equation~\ref{eqn:rb-ito-sys-net}, but with Gaussian noise added to the dictionaries at each instant in time. We defined $U(t):= U + N_t$ and $U_h(t):=U_h + N_t^h$ with $\text{vec}(N_t)\sim\cN\bigl(0, \sigma_N^2I\bigr)$ and $\text{vec}(N_t^h)\sim\cN\bigl(0, \sigma_{N_h}^2I\bigr)$ for all $t$. In this case, the relevant quantities $\Sigma_N, \Sigma_{N_h}$ appearing in Equation~\eqref{eqn:rb-avg-sys} take on the values $\Sigma_N=\dim(\cI)\sigma_N^2 I\in\bbR^{\dim(r)\times\dim(r)}$ and $\Sigma_{N_h}=\dim(r)\sigma_{N_h}^2 I\in\bbR^{\dim(z)\times\dim(z)}$. We assume that when there is more than one copy of the system, each copy  carries out coding with respect to the same noisy dictionaries -- the noise affecting the dictionaries is the same across copies. For these experiments, the Rao-Ballard hierarchies, network topology, and simulation parameters are identical to those described in Section~\ref{sec:expt-rb-error-signal}, unless otherwise noted.

\begin{figure}[p]
\centering
\includegraphics[width=0.9\textwidth]{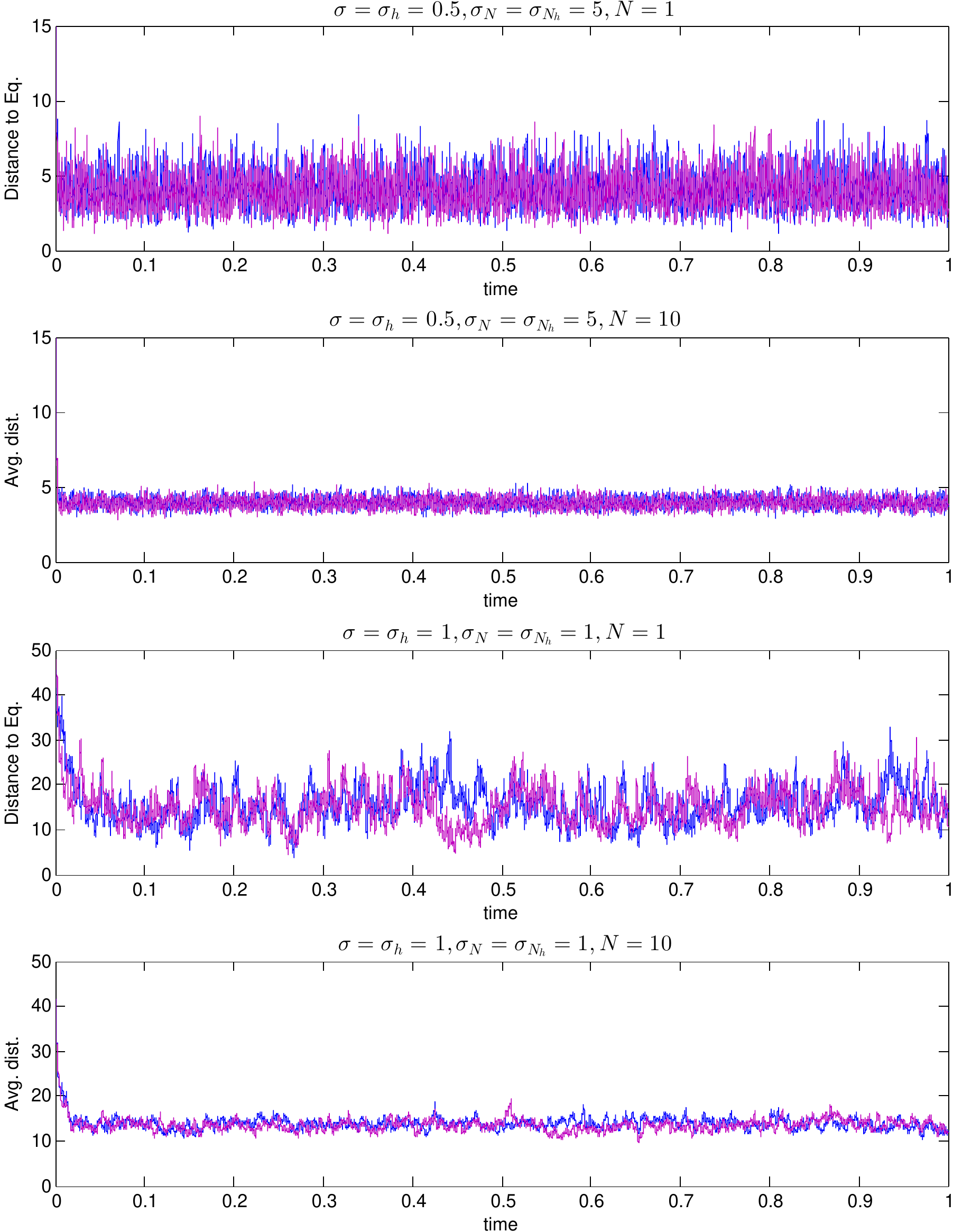}
\caption{\small Simulation of synchronized two layer Rao-Ballard networks with noisy top-down/bottom-up error signals and noisy dictionaries. The plots show distance from the mean-field trajectory to the expected equilibrium for various noise amplitudes and network sizes. See text for details. \textcolor{violet}{\bf Purple} traces indicate trajectories of the averaged system~\eqref{eqn:rb-avg-sys} while \textcolor{blue}{\bf blue} traces correspond to the original system. }
\label{fig:rb-dictnoise-network}
\end{figure}

Figure~\ref{fig:rb-dictnoise} gives distance of the system's trajectory to the expected equilibrium 
\[
\mu = (A + D)^{-1}c
\]
for three simulations of a single system with noisy dictionaries in which the dictionary noise variance was set to  $\sigma_N=\sigma_{N_h}=0.1, \sigma_N=\sigma_{N_h}=1$, and $\sigma_N=\sigma_{N_h}=5$ (in top-to-bottom order). The quantities $A,D$ and $c$ are as defined in Section~\ref{sec:rao-bal-dictnoise}. For each simulation the error signal noise variance was kept fixed at $\sigma=\sigma_{h}=0.5$. Blue traces indicate the distance to equilibrium of the original system, while purple traces give the distance to equilibrium of the averaged approximation~\eqref{eqn:rb-avg-sys}. As expected, noise on the dictionary can be seen to improve the convergence rate markedly, but it cannot be increased without also incurring additional noise variance at equilibrium (see bottom plot in particular). 

In Figure~\ref{fig:rb-dictnoise-network}, we show that synchronizing multiple systems reduces noise variance at equilibrium arising from both the error signal and dictionary noise sources. The top plot in Figure~\ref{fig:rb-dictnoise-network} is a rescaled version of the bottom plot in Figure~\ref{fig:rb-dictnoise}, where there is considerable noise amplitude at equilibrium due to the large additive noise on the dictionaries ($\sigma_N=\sigma_{N_h}=5$). In Figure~\ref{fig:rb-dictnoise-network}, second plot down, we show a simulation with the same noise amplitudes but with $N=10$ coupled copies of the system. The mean-field trajectories of both the original network (blue trace) and a network of averaged systems (purple trace) are shown. In the bottom two plots of Figure~\ref{fig:rb-dictnoise-network}, we again compare a single system to a network of coupled copies, but where the dictionary noise is on par with the error-signal noise and does not dominate the convergence rate or variance at equilibrium. In this case the network mean-field trajectory has considerably lower variance around equilibrium than the trajectory of a single hierarchy.

In all simulations, the approximate, averaged system can be seen to accurately capture the original system. From these plots, one can also deduce (in addition to doing so numerically) that the noisy system converges in expectation to the same expected equilibrium $\mu$ as the averaged system, confirming that the noisy systems indeed converge to the solution $\mu = (A+D)^{-1}c$ of the {\em regularized} optimization problem.

\subsection{PSGD with an objective exhibiting multiple local minima}
\begin{figure}[pt]
\centering
\includegraphics[width=0.48\textwidth]{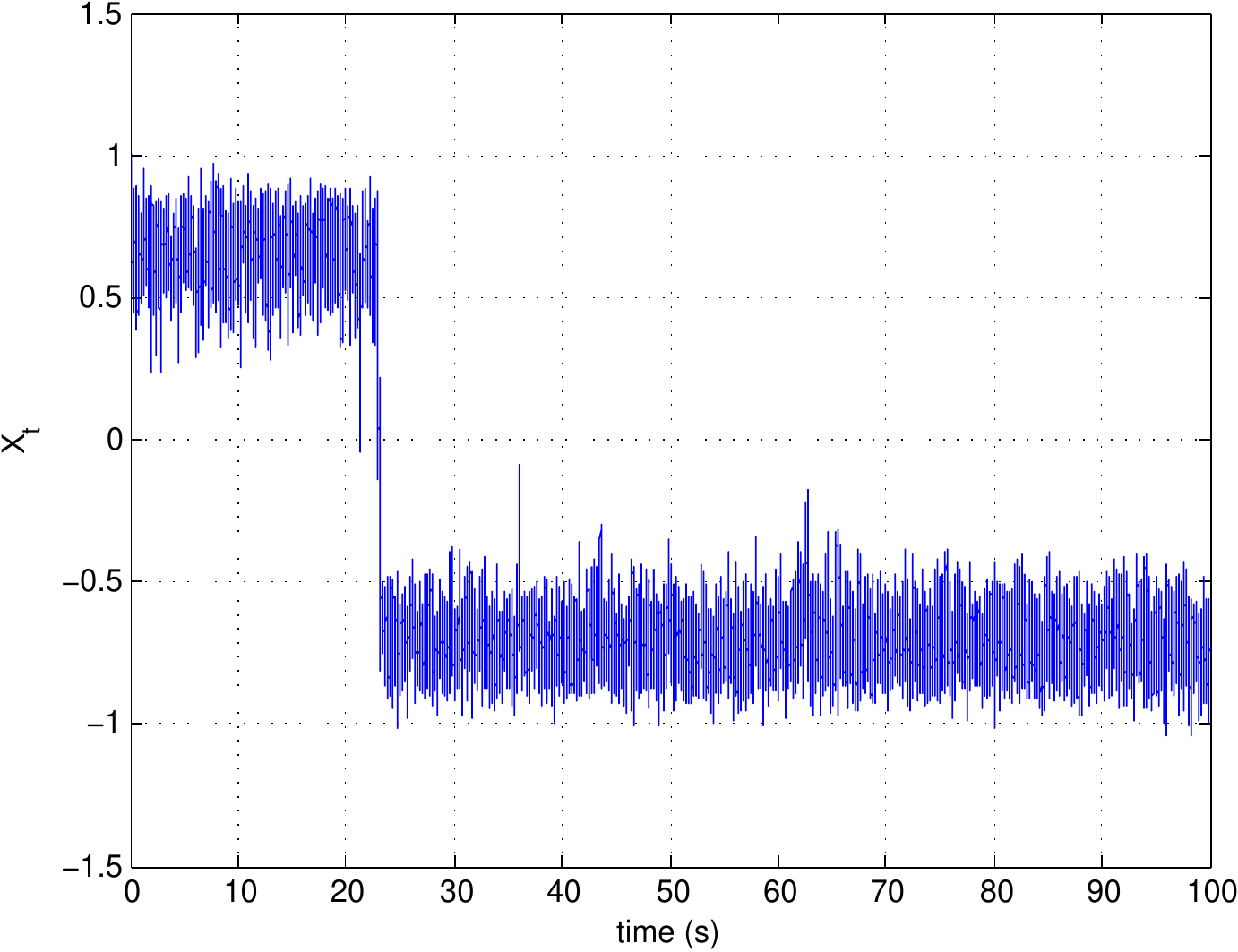}
\hspace{0.1cm}
\includegraphics[width=0.48\textwidth]{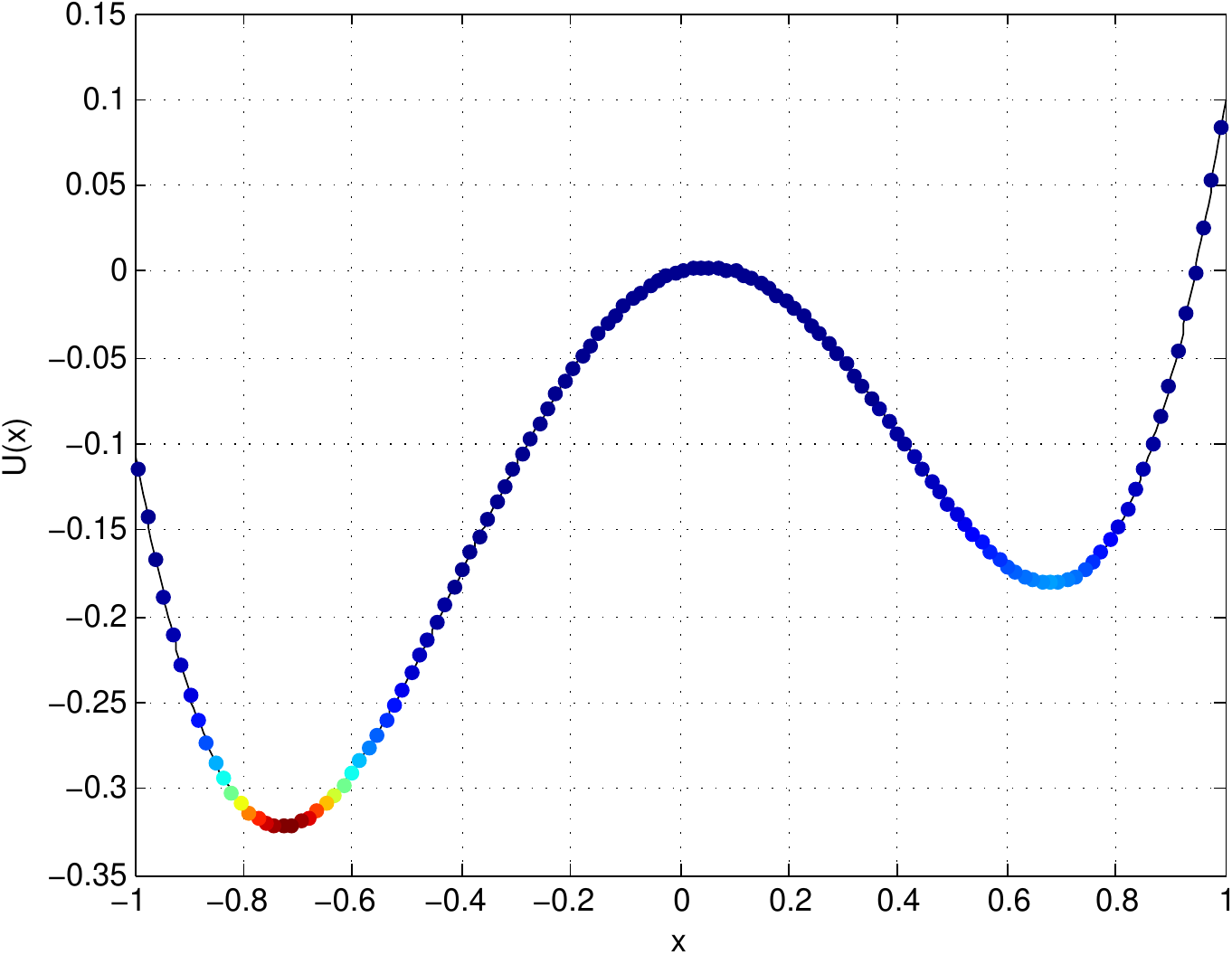} \\
\vspace{0.2cm}
\includegraphics[width=0.48\textwidth]{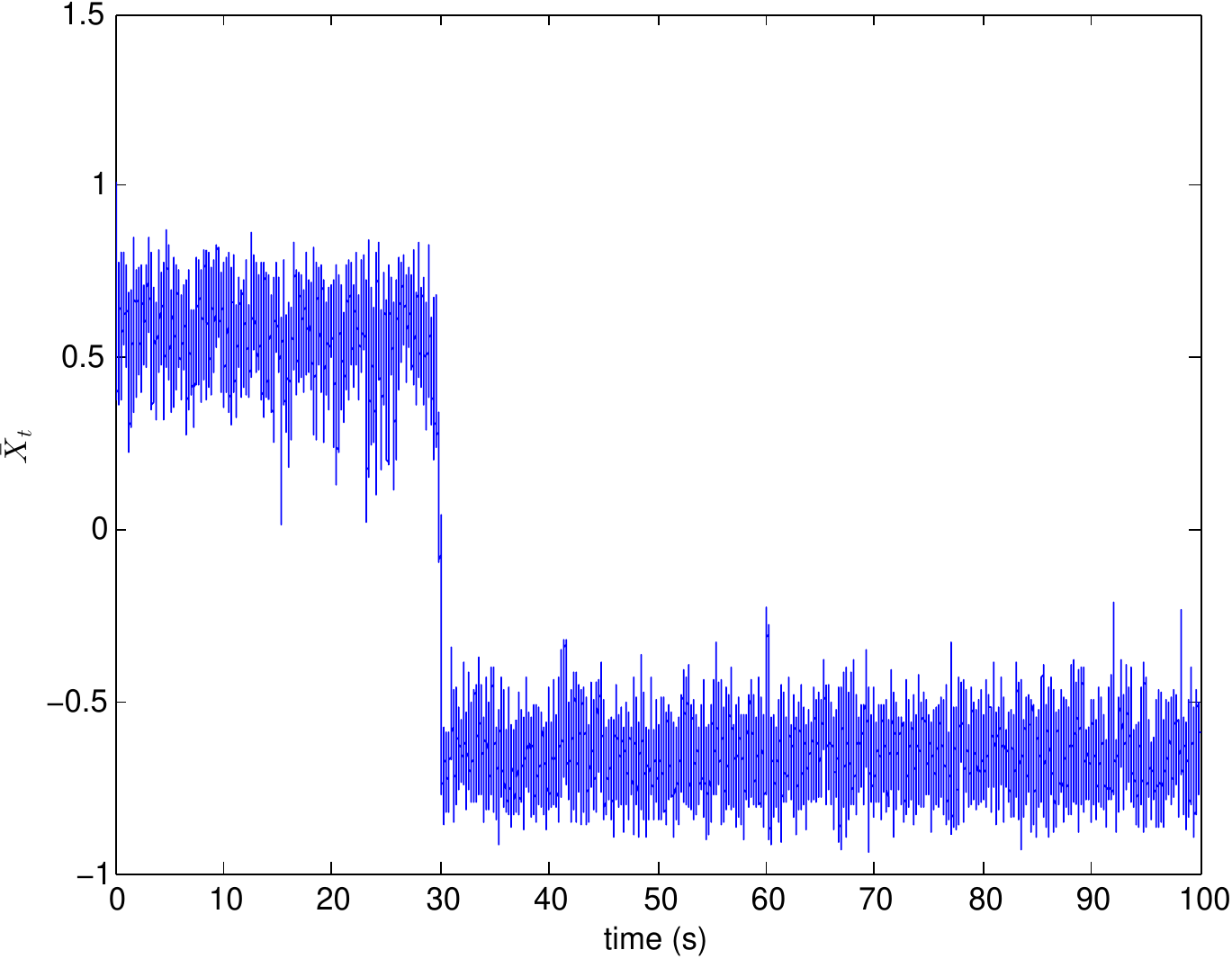}
\hspace{0.1cm}
\includegraphics[width=0.48\textwidth]{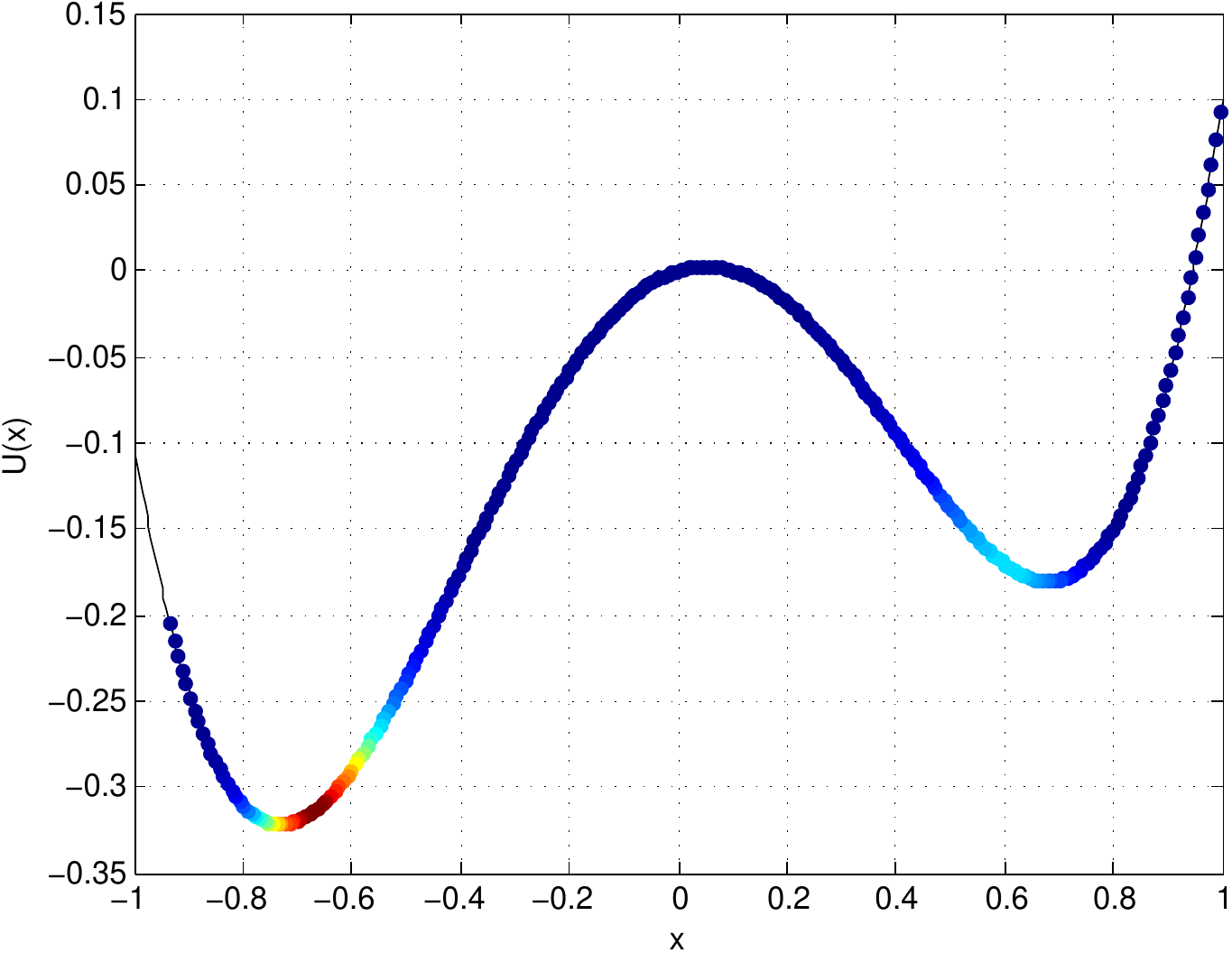} \\
\vspace{0.2cm}
\includegraphics[width=0.48\textwidth]{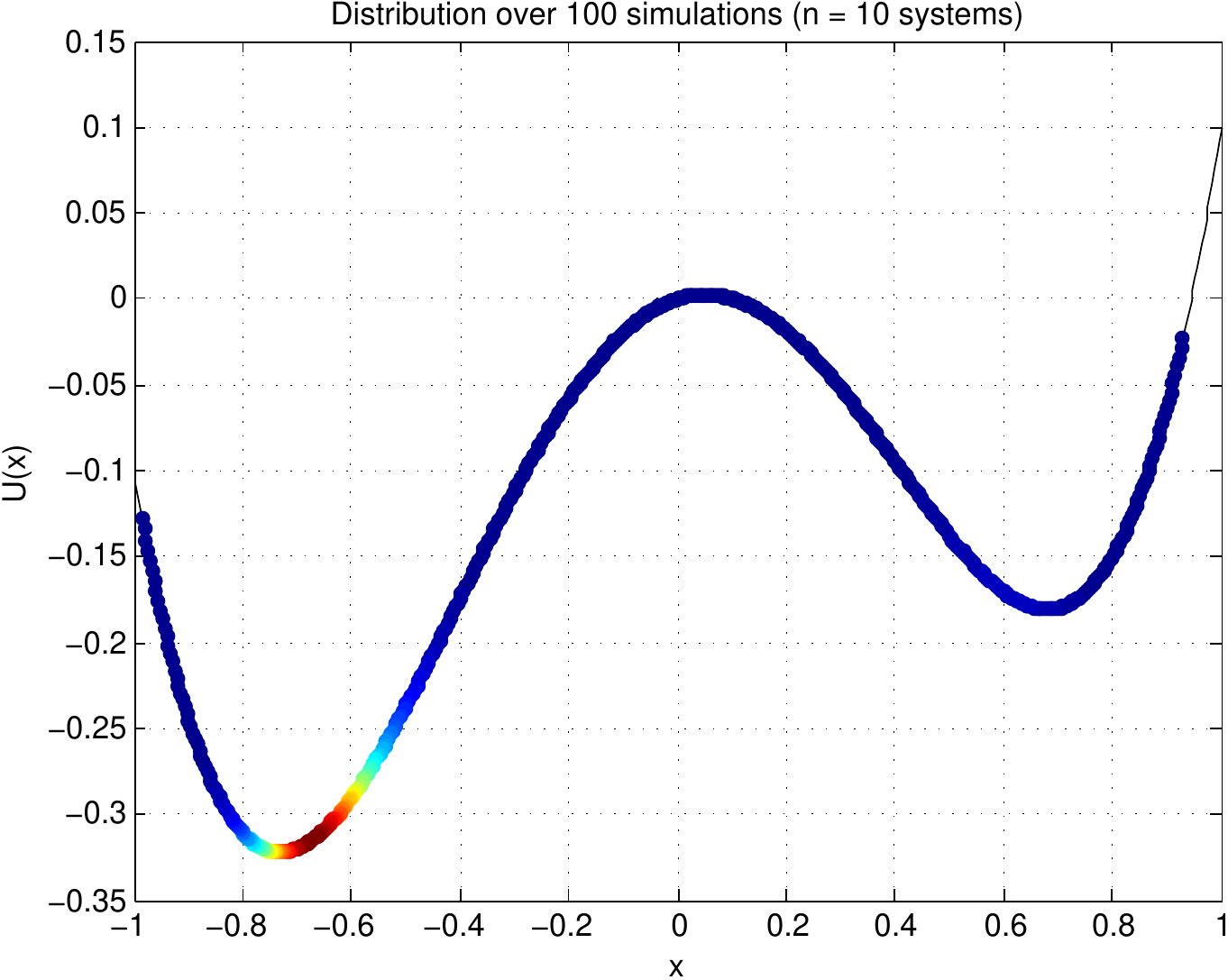}
\caption{\small PSGD simulations involving an objective with two local minimums and noisy dynamics. Top row: $n=1$ system. Middle row: $n=10$ coupled systems, all with initial conditions at $x=1$. Left-hand plots give the average trajectory, and right-hand plots illustrate empirical state distributions as heat maps superimposed on the objective. Bottom row: state distribution over 100 simulations with $n=10$ and initial conditions drawn i.i.d. from the uniform distribution on $[-1,1]$.}
\label{fig:spsa-potential}
\end{figure}

To illustrate PSGD in the context of a nonlinear, non-convex objective function, we simulated 
a network of coupled diffusions in a double-well potential using the PSGD gradient as an approximation to the true gradient. Because the dynamics are nonlinear, synchronization is necessary to average out the effects of the noise and to obtain the consensus trajectory without global pooling. We considered the one-dimensional objective $$U(x)= x^4 -x^2 +0.1x$$ the shape of which may be seen in Figure~\ref{fig:spsa-potential} (right column of plots). For simplicity, we computed the PSGD gradient with Gaussian perturbations $\xi_t$, rather than with an OU approximation. The collection of $n$ identical but independent 1-D SDEs were coupled with diffusive coupling according to an all-to-all network topology with uniform strength $\kappa$, giving the system
\begin{equation}
dX_t = -\frac{\gamma}{\sigma_N^2}\bigl[\bU(X_t+\xi_t)-\bU(X_t)\bigr]\xi_tdt + LX_t + \sigma^2dB_t
\end{equation}
where $X_t=(X_t^1,\ldots,X_t^n)^{\tr}, \bU(X)=(U(X^1),\ldots,U(X^n))^{\tr}, \xi_t\sim\cN(0,\sigma_N^2 I), \forall t$, $B_t$ is standard $n$-dimensional Brownian motion independent of $\xi$, and $L$ is the network Laplacian.

In the simulations that follow, we selected $\sigma_N=0.05$ (small enough so that the Taylor approximation beneath the PSGD scheme is reasonable), $\gamma=10$ and $\kappa=4$. The latter two parameters were chosen to give reasonable convergence and synchronization rates. The integration stepsize was chosen  $\Delta t=10^{-4}$ uniformly over the simulation interval $0\leq t \leq 100s$.

The top row of Figure~\ref{fig:spsa-potential} shows a simulation consisting of only one SDE ($n=1$) starting from $X(0)=1$ with $\sigma=0.8$. This value of $\sigma$ is just large enough to see switching between equilibria with high probability inside the simulation time interval. The system's trajectory is shown on the left, while the empirical state density is depicted as a heat map plotted along the objective function on the right. Red colors indicate that the system spent large amounts of time in or near the corresponding $x$-coordinates on the plot, while blue colors mean that the system spent little time in the corresponding $x$ states. Around $t=30s$, there is a clear transition from the local minimum near $x=0.7$ to the global minimum near $x=-0.7$.

The second row of plots in Figure~\ref{fig:spsa-potential} summarizes a similar simulation, but with $n=10$ coupled SDEs and $\sigma=2$. This value for for $\sigma$ was again chosen just large enough to observe switching between equilibria in the mean-field trajectory with high probability. Here, the left plot shows the center-of-mass trajectory $\bar{X}_t=\frac{1}{n}\sum_i X_t^i$ assuming $X^i(0)=1, \forall i$, and the right plot provides a visualization of the estimated state density for $\bar{X}_t$. Placing initial conditions at a worst-case $x=1$ for every diffusion all but forces the overall system to either stay in the suboptimal local minimum, or switch back and forth between the two stable minima. The average trajectory exhibits a fast transition between equilibria, and stays close to those equilibria, precisely because the individual trajectories are close to each other and maintain this closeness with a comparatively fast synchronization rate (plots of the individual trajectories differ negligibly). 

The bottom plot in Figure~\ref{fig:spsa-potential} shows the empirical state distribution computed from 100 simulations identical to the example with $n=10$ SDEs immediately above, but where the initial conditions for each simulation were drawn independently from the uniform distribution on $[-1,1]^n$. This scenario represents a more practical setting, since one usually does not know in advance where the local minima are, and the best we can do is try to reach a good one from a random starting point. Ideally, synchronizing multiple randomly initialized PSGD systems might lead to a consensus solution corresponding to a decent local solution. The distribution confirms that for this simple problem the system does consistently find the better of the two minima.

\begin{figure}[tp]
\centering
\includegraphics[width=0.48\textwidth]{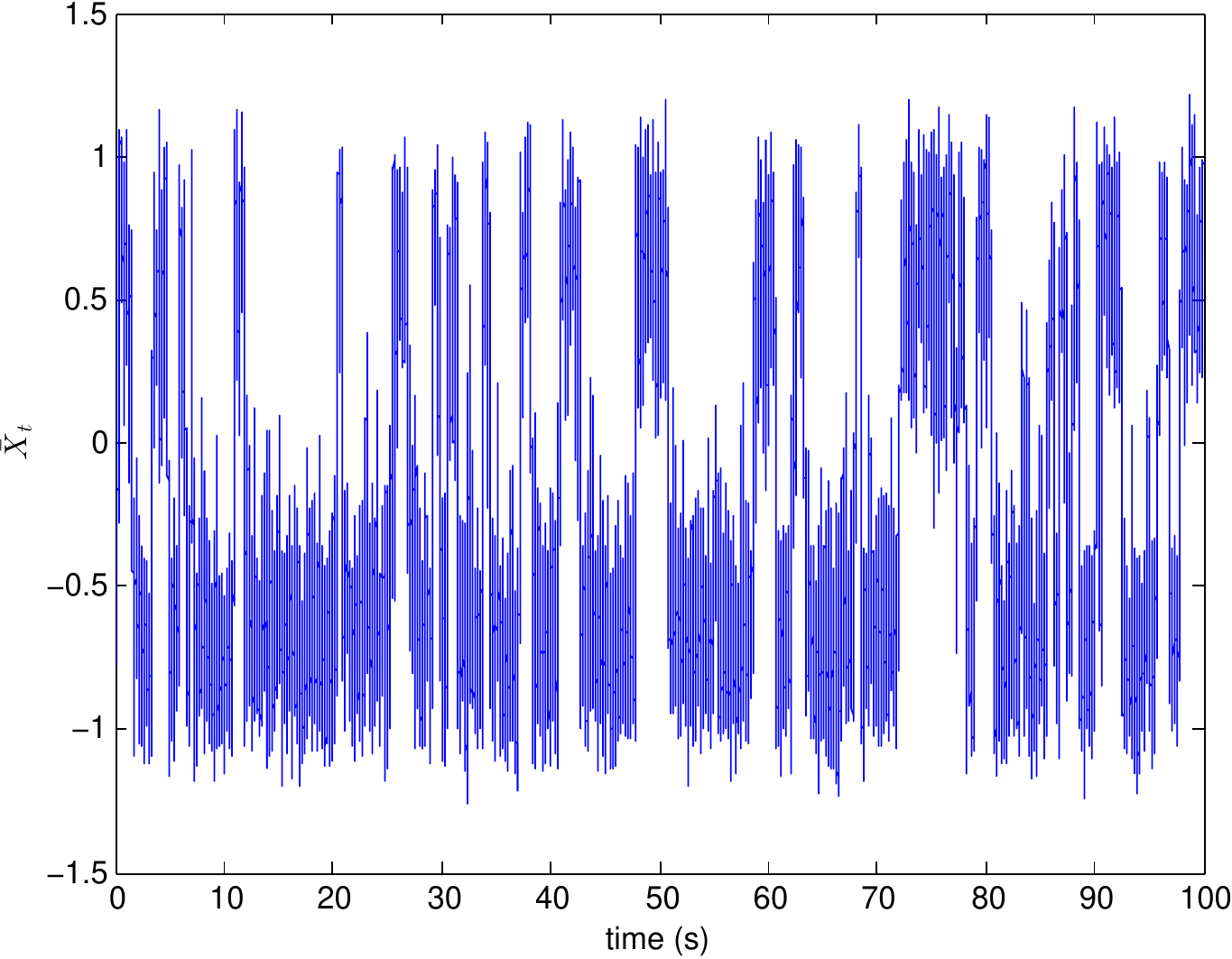}
\hspace{0.1cm}
\includegraphics[width=0.48\textwidth]{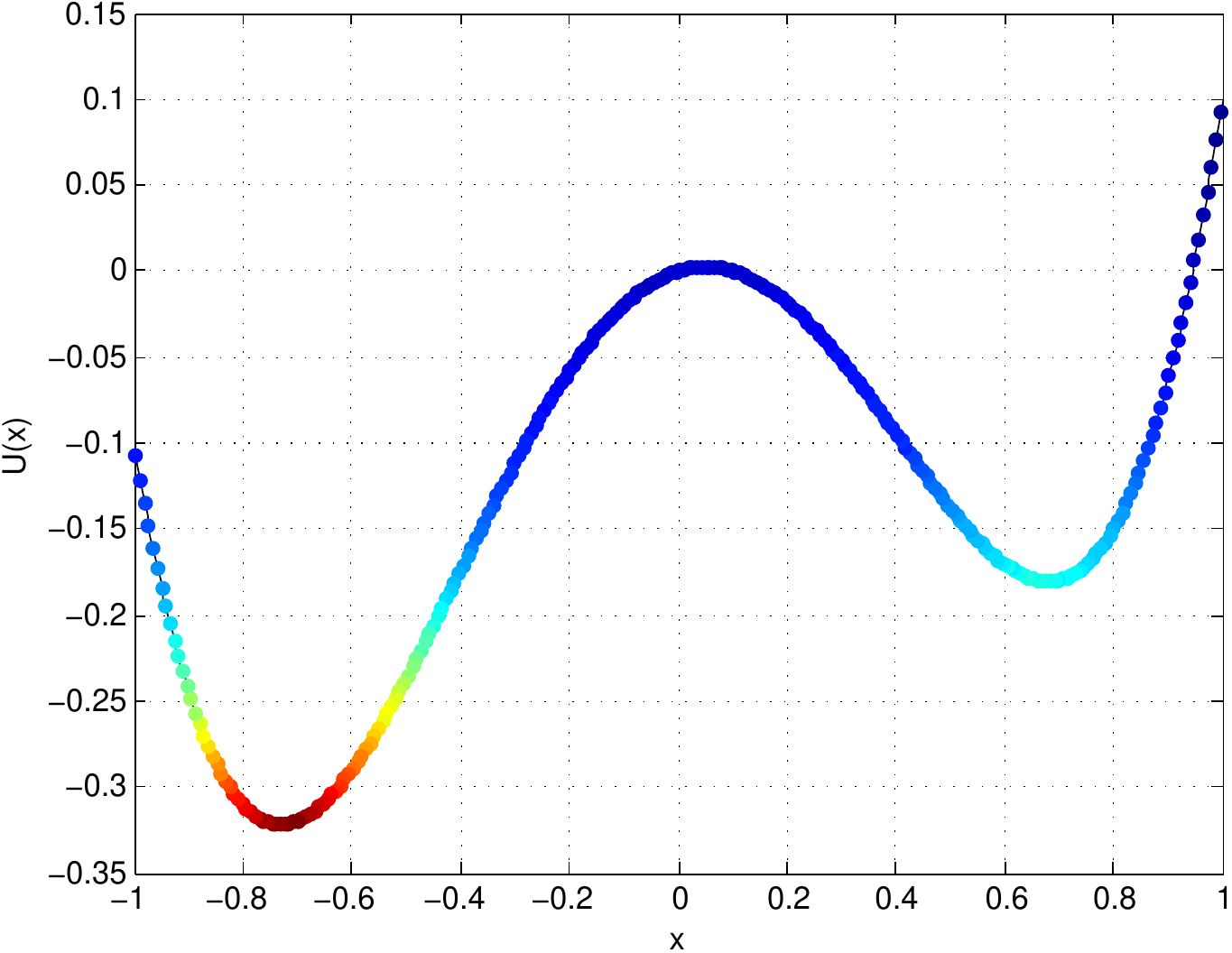} \\
\vspace{0.2cm}
\includegraphics[width=0.48\textwidth]{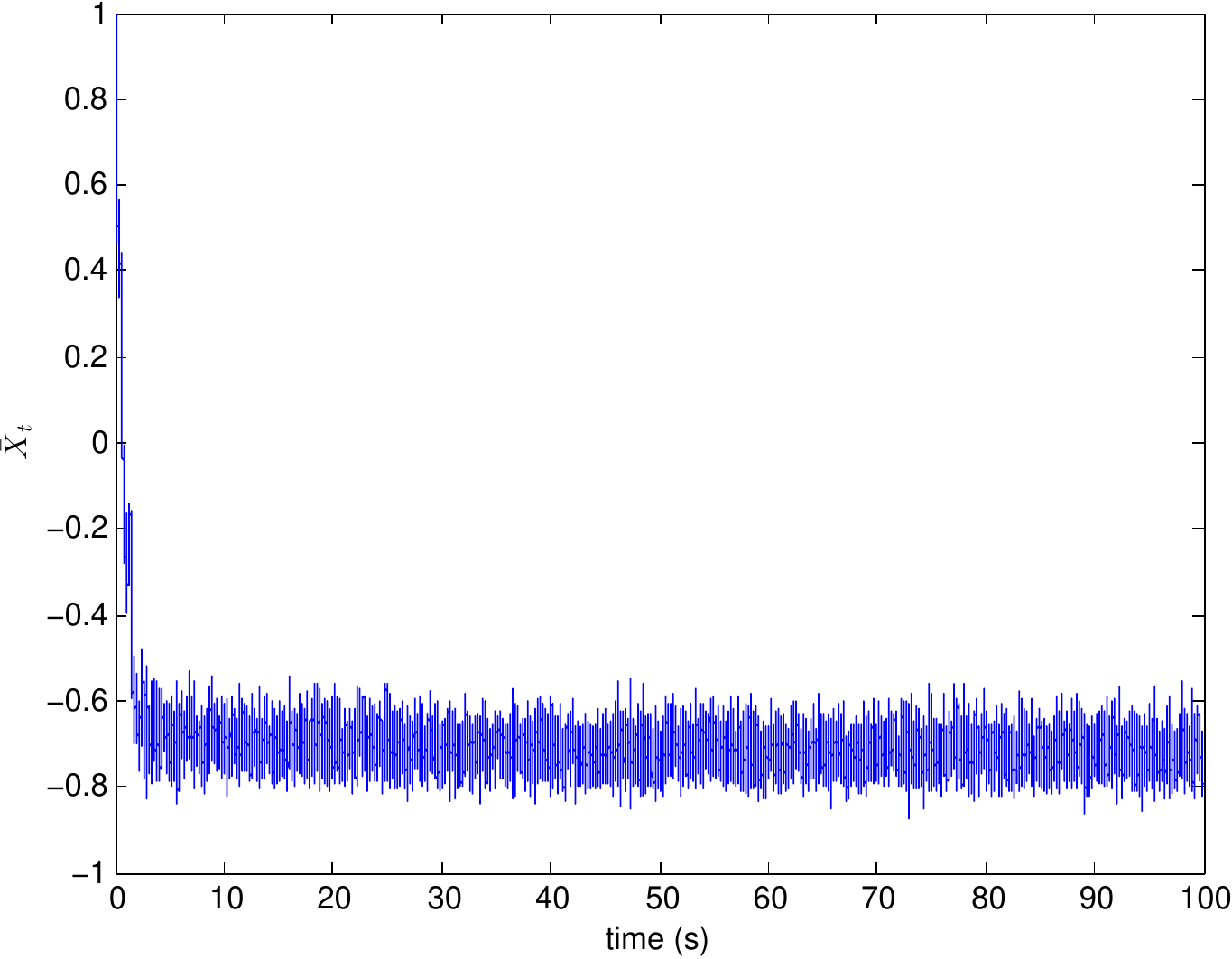}
\hspace{0.1cm}
\includegraphics[width=0.48\textwidth]{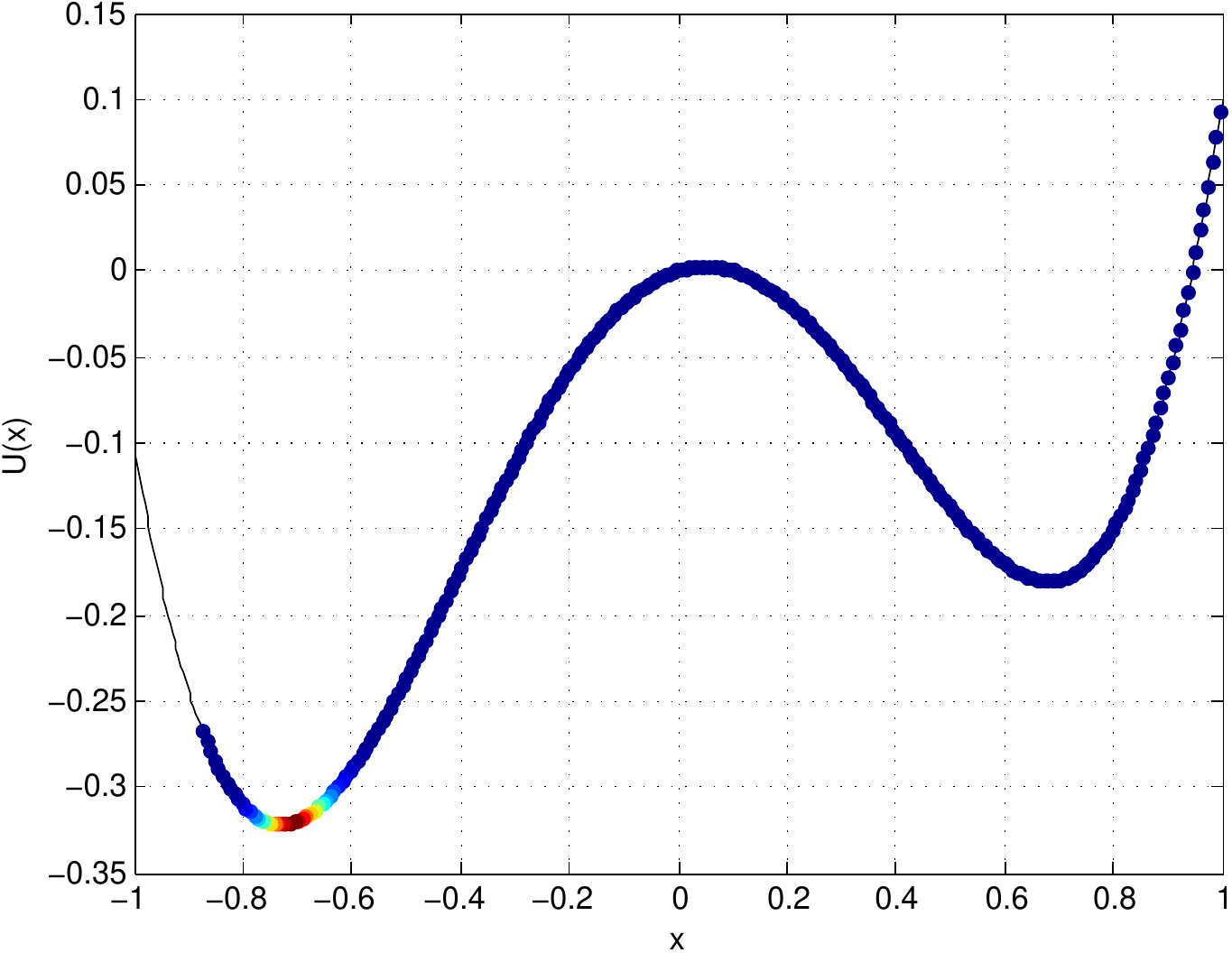}
\caption{\small A network of synchronized PSGD dynamical systems can be used to prevent switching between minima. Here the diffusion noise is large, with $\sigma=1.6$. Top row: $n=1$ system, $X(0)=1$. Second row: $n=20$ coupled systems, all with $X^i(0)=1$, and with time-varying coupling strength $\kappa(t)=20(t/100)$. Left-hand plots show the average trajectory, and right-hand plots illustrate empirical state distributions as heat maps superimposed on the objective.}
\label{fig:spsa-switching}
\end{figure}

The last pair of simulations, shown in Figure~\ref{fig:spsa-switching}, demonstrate that a network of coupled systems can be used to prevent switching back and forth between equilibria, thereby improving stability of the consensus solution. We consider the setting in which the noise level may not be under direct control of the agent or user, and is kept fixed at $\sigma=1.6$. This noise deviation is large enough to cause any one PSGD system to repeatedly switch between the two minima over time, as may be seen in the top row of plots in  Figure~\ref{fig:spsa-switching} where a single system has been simulated. 

To encourage a collection of coupled systems to find the global minimum, and then remain there with high probability, we set $n=20$ systems and  allowed the coupling strength to gradually increase over time from $\kappa=0$ at $t=0$ to $\kappa=20$ at $t=100$ according to a linear schedule. Here, we might interpret the coupling strength as a measure of ``confidence''. Low initial confidence allows enough noise variance initially to search out multiple solutions. However, as confidence increases, the entire system is kept in a neighborhood of the consensus equilibrium: if any one trajectory momentarily jumps out of the minimum, the other elements quickly pull it back due to the strong coupling. At the noise amplitude and coupling strength chosen above, we empirically observed that multiple trajectories will make large jumps nearly simultaneously only with very low probability. Thus, the probability that the consensus jumps is low. The result of this simulation, again assuming $X^i(0)=1, \forall i$, is shown in the bottom row of plots in Figure~\ref{fig:spsa-switching}. The path of the average trajectory shows that the system quickly finds the global minimum and remains there for the duration of the simulation. As expected, the variance around this minimum is also seen to be reduced compared to the simulation of one system.


\subsection*{Acknowledgments}
The authors are grateful to Rodolfo Llinas for pointing out the plausible analogy between gradient search in adaptive optics and learning mechanisms in the brain. Some of the work described in this paper was completed while JB was with the Department of Mathematics at Duke University, supported under contracts DARPA  FA8650-11-1-7150 SUB\#7-3130298, NSF IIS-08-03293 and WA State U. SUB\#113054 G002745, all to M. Maggioni.

\appendix
\section{Proofs}
\subsection{Proof of Theorem~\ref{thm:sys-convergence}}
Recall the expressions for $\bmu_w(t)$ and $\Sigma_w(t)$ given in Section~\ref{sec:noise-gives-reg},  Equation~\eqref{eqn:w-mean-process}, and let $\mu$ denote the common (noise-free) equilibrium value. Viewing $\bw(t)$ as a collection of Gaussian random variables indexed by $t$, expressions for $\bar{w}(t)$ and $\widetilde{\bw}(t)$ can be obtained as manipulations of Gaussians:
\begin{align*}
\bar{w}(t) &\sim \cN\bigl(\tfrac{1}{n}\bbone^{\tr}\bmu_w(t),\, \tfrac{1}{n^2}\bbone^{\tr}\Sigma_w(t)\bbone\bigr)\\
&=\cN\bigl(e^{-\alpha t}\bbE[\bar{w}(0)] + (1-e^{-\alpha t})\mu,\,
e^{-2\alpha t}\bbE[(\bar{w}(0))^2] + \tfrac{\sigma^2}{2\alpha n}(1 - e^{-2\alpha t})
\bigr)
\end{align*}
where $\bar{w}(0)=\tfrac{1}{n}\bbone^{\tr}\bw(0)$. Turning to the fluctuations, let $Q=I-\tfrac{1}{n}\bbone\bbone^{\tr}$ denote the orthogonal projection onto the zero-mean subspace of $\bbR^n$. Note that $\widetilde{\bw}=Q\bw$ and $\bar{w}\bbone = (I-Q)\bw$. We have
\begin{align*}
\widetilde{\bw}(t) &\sim \cN(Q\bmu_w(t),\, Q\Sigma_w(t)Q^{\tr}) \\
&= \cN\bigl( e^{-(L + \alpha I)t}\bbE[\widetilde{\bw}(0)],\,
 e^{-(L+\alpha I)t}\bbE[\widetilde{\bw}(0)\widetilde{\bw}(0)^{\tr}]e^{-(L+\alpha I)t} \\
 & \hskip 3.75cm + \tfrac{\sigma^2}{2}(QLQ^{\tr} + \alpha I)^{-1}(I - e^{-2(QLQ^{\tr} + \alpha I)t})
\bigr) .
\end{align*}

We can now consider the error
\[
\bbE\bigl[\tfrac{1}{n}\nor{\bw(t) - \bmu}^2\bigr] =
\bbE\bigl[\tfrac{1}{n}\nor{\widetilde{\bw}(t)}^2\bigr] +
\bbE\bigl[\tfrac{1}{n}\nor{\bar{w}(t)\bbone - \bmu}^2\bigr] .
\]
 In general if $\bx\sim\cN(\bmu_x,\Sigma_x)$ then $\bbE[\nor{\bx-\bc}^2] = \trace(\Sigma_x) + \nor{\bmu_x-\bc}^2$ for any (non-random) vector $\bc$.
The first error term on the right-hand side can be estimated as
\begin{alignat*}{2}
\bbE\bigl[\tfrac{1}{n}\nor{\widetilde{\bw}(t)}^2\bigr] &\leq
\begin{aligned}[t]
 \frac{1}{n}\sum_{i>0} \lambda_i(\Sigma_w(0)) e^{-2(\lambda_i(L) + \alpha)t} &+
\frac{\sigma^2}{2n}\sum_{i>0}\frac{1-e^{-2(\lambda_i(L) + \alpha)t}}{\lambda_i(L)+\alpha} \\
  &+ \bbE[\bw(0)]^{\tr}Q^{\tr}e^{-2(L + \alpha I)t}Q\bbE[\bw(0)]
\end{aligned} \\
 &\leq \lambda_{\text{max}}(\Sigma_w(0)) e^{-2(\lmin + \alpha)t} +
\frac{\sigma^2\bigl(1-e^{-2(\lmin + \alpha)t}\bigr)}{2(\lmin+\alpha)} +
e^{-2(\lmin + \alpha)t}\nor{\bbE[\bw(0)]}^2
\end{alignat*}
where $\lmin$ is the smallest non-zero eigenvalue of $L$ and $\lambda_{\text{max}}(\cdot)$ denotes the largest eigenvalue of its argument. The first term on the right-hand side of the first inequality follows from Von Neumann's trace inequality. The second error term is given by
\begin{align*}
\bbE\bigl[\tfrac{1}{n}\nor{\bar{w}(t)\bbone - \bmu}^2\bigr] &= \bbE\bigl[\bigl(\bar{w}(t) - \mu\bigr)^2\bigl] \\
& = e^{-2\alpha t}\bbE[(\bar{w}(0))^2] + \frac{\sigma^2}{2\alpha n}(1 - e^{-2\alpha t}) +
e^{-2\alpha t}\bigl(\bbE[\bar{w}(0)] - \mu\bigr)^2.
\end{align*}
Note that $\bbE[(\bar{w}(0))^2] = \tfrac{1}{n^2}\bbone^{\tr}\Sigma_w(0)\bbone$ and
$\bbE[\bar{w}(0)]=\tfrac{1}{n^2}\bbone^{\tr}\bmu_w(0)$.
Defining the constants
\begin{align*}
\widetilde{C} &:= \lambda_{\text{max}}(\Sigma_w(0)) - \frac{\sigma^2}{2(\lmin+\alpha)} +
\nor{\bbE[\bw(0)]}^2 \\
\overline{C} &:= \bbE[(\bar{w}(0))^2] - \frac{\sigma^2}{2\alpha n} +
\bigl(\bbE[\bar{w}(0)] - \mu\bigr)^2
\end{align*}
and combining with the above, we obtain the Theorem. $\qed$

\subsection{Proof of Theorem~\ref{thm:ou-sync-noisered}}
If the initial condition is deterministic or Gaussian distributed, any solution to the system is a Gaussian (Ornstein-Uhlenbeck) process. The stationary distribution of the system is the distribution of $X_{\infty}$, and has mean $\mu$ by inspection of the SDE. The long-term covariance is given by 
\[
\text{cov}(X_{\infty}) = \bbE\bigl[(X_{\infty}- \bbone_M\otimes\mu)(X_{\infty}-\bbone_M\otimes\mu)^{\tr}\bigr] 
= \lim_{t\to\infty}\int_0^t e^{(L\oplus A)(s-t)}(I_M \otimes \Sigma\Sigma^{\tr}) e^{(L\oplus A)^{\tr}(s-t)}ds ,
\]
assuming that $(L\oplus A)$ is positive (semi)-definite so that the integral converges. 
Since both $L$ and $A$ are symmetric, we can expand $L\oplus A$ in terms of its orthonormal eigenbasis, 
\mbox{$
L\oplus A = \sum_{i=1}^{Md}\lambda_iP_i,
$}
where $P_i$ is the rank one orthogonal projector onto the eigenspace associated to eigenvalue $\lambda_i$. Substituting this expansion into the integral above gives,
\begin{align*}
\text{cov}(X_{\infty}) &= \sum_{i,j}P_i(I_M \otimes \Sigma\Sigma^{\tr})P_j
\left[\lim_{t\to\infty}\int_0^t e^{(\lambda_i+\lambda_j)(s-t)}ds\right] \\
&=  \sum_{i,j}\frac{1}{\lambda_i+\lambda_j}P_i(I_M \otimes \Sigma\Sigma^{\tr})P_j .
\end{align*}
Since $\bbE[\nor{X_{\infty} - \bbone_M\otimes\mu}^2] = \trace[\text{cov}(X_{\infty})]$, we only actually need the trace of each term above. This is helpful because, for any $M$ of appropriate size, $\trace(P_iMP_j)=0, \forall i\neq j$. Applying this observation to the above, we have
\begin{align*}
2\trace\bigl[\text{cov}(X_{\infty})\bigr] &= \sum_i\frac{1}{\lambda_i}\trace\bigl[P_i(I_M \otimes \Sigma\Sigma^{\tr})\bigr] \\
&\leq \sum_i\frac{\lambda_{\text{max}}(\Sigma\Sigma^{\tr})}{\lambda_i} \\
&\leq \lambda_{\text{max}}(\Sigma\Sigma^{\tr})\left(\frac{1}{\lambda_{\text{min}}(A)} + \frac{Md}{\lambda_{\text{min}}(A) + \lmin}\right).
\end{align*}
The final inequality follows from the fact that the eigenvalues of $L\oplus A$ are of the form $\lambda_i(L)+\lambda_j(A)$ for $i,j=1,\ldots,d$. Since $\lambda_1(L)=0$, $\lambda_{\text{min}}(A)$ is the smallest eigenvalue of $L\oplus A$, and $\lambda_{\text{min}}(A) + \lmin$ is the second smallest eigenvalue. Rewriting $\nor{X_{\infty} - \bbone_M\otimes\mu}^2$ as $\sum_{i=1}^M\nor{X_{\infty}^i-\mu}^2$  on the left-hand side and dividing through by $M$ completes the Theorem.$\qed$

\subsection{Proof of Theorem~\ref{thm:quad-pgsd-sys}}
The OU process~\eqref{eqn:noise_proc} is ergodic and has stationary distribution 
$\mu_{\infty}=\cN(\mathbf{0}, \tfrac{1}{2}\sigma^2 I)$. Furthermore, the system \eqref{eqn:quad_dyn}-\eqref{eqn:noise_proc} satisfies the conditions of Theorem~\ref{thm:averaging}.  Homogenizing~\eqref{eqn:quad_dyn} requires the averaged vector field
\begin{equation*}
F(u) = \int_{\bbR^d}(2u^{\tr}Az + z^{\tr}Az)z\mui(\dif z)
 =2\bbE[zz^{\tr}]Au + \bbE[zz^{\tr}Az]
 =\sigma^2Au
\end{equation*}
(using that odd moments of a zero-mean Gaussian are zero), and leads to the averaged system
$$
\dot{U} = -\gamma\sigma^2AU,\qquad U(0)=\bu(0).
$$
The solution to this ODE is easily found to be
$
U(t) = e^{-\gamma\sigma^2At}U(0) .
$
Theorem~\ref{thm:averaging} then provides that $\bu(t)$ converges in distribution to $U(t)$ as $\varepsilon\to 0$. Since $U(t)$ is deterministic for all $t\geq 0$ in this case, $\bu(t)$ also converges to $U(t)$ in probability. Let $(e_i)_{i=1}^d$ denote the canonical basis of $\bbR^d$ and let $x^i$ denote the $i$-th coordinate of a vector $x\in\bbR^d$. The projection function $\pi_i(x)=\scal{x}{e_i}=x^i$ is clearly continuous, so by the continuous mapping theorem, $\pi_i(\bu_t)\to U^i(t)$ in probability. 

Let $\bue(t)$  denote the (strong) solution to~\eqref{eqn:quad_dyn} for some fixed $\varepsilon\in(0,1]$.
If the family $\{\bue^i(t)\}_{\varepsilon\in(0,1]}$ is uniformly integrable (for each $t<\infty$), then together with $\bu^i(t)\to U^i(t)$ i.p., we would have that $\bbE[\bue^i(t)]\to\bbE[U^i(t)]=U^i(t)$ as $\varepsilon\to 0$ (by way of convergence in $L_1$). We establish uniform integrability by showing that $\sup_{\varepsilon\in(0,1]}\bbE[\pi_i^2\bigl(\bue(t)\bigr)] <\infty$. First note that for any $\varepsilon> 0$, the OU process~\eqref{eqn:noise_proc} is a Gaussian process $\bZ_t\sim\cN(\bmu_t,\Sigma_t)$ with bounded moments $\bbE[\nor{\bZ_t}^p]<\infty, p\geq 1$, for all $t\leq T<\infty$: Suppose $X\sim\cN(0,I_{d\times d})$. Then for each $t$, $\bZ_t = \bmu_t + \Sigma_t^{1/2}X$ in law. Because the standard Normal moments $\bbE[\nor{X}^p]$ are bounded for all $p$, we have, restricting our attention to $p$ even, that 
\begin{align*}
\bbE[\nor{\bZ_t}^p] &\leq 2^{p-1}\bigl(\nor{\bmu_t}^p + \bbE[\nor{\Sigma_t^{1/2}X}^p]\bigr)\notag\\
&\leq C_p\bigl(e^{-pt/\varepsilon} + (\trace\Sigma_t)^{p/2}\bbE[\nor{X}^p]\bigr)\notag\\
&\leq C_p(1 + e^{-pt/\varepsilon}) < \infty \label{eqn:Z-moments}
\end{align*}
where $C_p$ is a constant depending on $p$ that changes from instance to instance, and where $\bmu_t,\Sigma_t$ are given by~\eqref{eqn:ou-net-mu},~\eqref{eqn:ou-net-sigma} (resp.) with $L_z=0, \eta=1, \gamma=\sigma/\sqrt{2}$. Returning to the second moment of $\bu$, define the norm $\nor{x}_A\triangleq \sqrt{\scal{x}{Ax}}$, where $A$ is the symmetric strictly positive definite matrix appearing in~\eqref{eqn:quad_dyn}. Note that $\lambda_{\mathrm{min}}(A)\nor{x}^2\leq \nor{x}_A^2\leq \lambda_{\mathrm{max}}(A)\nor{x}^2$ for any $x\in\bbR^d$, where $\lambda_{\mathrm{min}}(A) > 0$ is the smallest eigenvalue of $A$ and $\lambda_{\mathrm{max}}(A) < \infty$ is the largest eigenvalue of $A$.
Applying Ito's lemma to the map $\bu\mapsto\nor{\bu}_A^2$, we have for any $\varepsilon > 0$ and $0 \leq t\leq T<\infty$,
\begin{align*}
\bbE\bigl[\nor{\bue(t)}^2_A\bigr] &= -2\gamma\int_0^t\bbE\left[\bigl(2\bue(s)^{\tr}A\bZ_s + \bZ_s^{\tr}A\bZ_s\bigr)\bigl(\bue(s)^{\tr}A\bZ_s\bigr)\right]\dif s + \nor{\bu(0)}^2_A  \\
& \leq -2\gamma\int_0^t\bbE\left[(\bZ_s^{\tr}A\bZ_s)(\bue(s)^{\tr}A\bZ_s)\right]\dif s + \nor{\bu(0)}^2_A  \\
& \leq 2\gamma\int_0^t\bbE\left[\nor{\bue(s)}_A\nor{\bZ_s}^3_A\right]\dif s + \nor{\bu(0)}^2_A  \\
& \leq C\int_0^t\bbE\left[\nor{\bue(s)}_A^2 + \nor{\bZ_s}^6_A\right]\dif s + \nor{\bu(0)}^2_A  \\
& \leq C\int_0^t\bbE\bigl[\nor{\bue(s)}_A^2\bigr]\dif s + C\bigl(t + \varepsilon(1-e^{-t/\varepsilon}\bigr) + \nor{\bu(0)}^2_A  \\
& \leq C\bigl(1 + \varepsilon + \nor{\bu(0)}^2_A\bigr)e^{Ct} < \infty
\end{align*}
where $C$ is a constant independent of $\varepsilon$ that changes from line to line.  The second inequality follows using that  $-(\bue^{\tr}A\bZ)^2\leq 0$, the third from Cauchy-Schwarz, and the fourth follows from Young's inequality. The fifth line follows from substituting and integrating the estimate for $\bbE[\nor{\bZ_t}^p]$ computed above, and the final line follows from an application of Gronwall's inequality.

Since $\bbE[\nor{\bue(t)}^2]\leq (1/\lambda_{\mathrm{min}}(A))\bbE[\nor{\bue(t)}^2_A]$, the coordinates of $\bue(t)$ individually have bounded second moments for all $\varepsilon > 0$, and
$\sup_{\varepsilon\in(0,1]}\bbE[\pi_i^2\bigl(\bue(t)\bigr)] 
\leq C(1 + \sup_{\varepsilon\in(0,1]} \varepsilon e^{Ct}) < \infty$ for all $i$.
Hence, $\bbE[\bue^i(t)]\to U^i(t)$ for each $i$, and so $\bbE[\bue(t)]\to U(t)$ as $\varepsilon\to 0$. $\qed$

\subsection{Proof of Proposition~\ref{prop:spsa-homogenization}}
We homogenize the entire system of $n$ SDEs by homogenizing each individual coordinate's dynamics separately.
For any $i\in\{1,\ldots,n\}$, we may write down a smaller system describing the evolution of $w_t^i$:
\begin{align*}
dx &= -(axy^2 + (L\bx)_i)dt + \frac{1}{\sqrt{\varepsilon}}\left(-\frac{a}{2}y^3 + z\right)dt \\
d\begin{pmatrix}
y \\ z
\end{pmatrix} 
&= \frac{1}{\varepsilon}\begin{pmatrix}
-y \\ -z
\end{pmatrix}dt
 + \frac{1}{\sqrt{\varepsilon}}\begin{pmatrix}
 \sqrt{2}\sigma_N & \\
  & \sqrt{2}\sigma_w
 \end{pmatrix}
 \dB_t
\end{align*}
where, in the interest of readability, we have adopted the simplified notation $x:=w_t^i, y:=N_t^i, z := U_t^i$ and $a:=\gamma\alpha, \bx:=\bw$.
The generator of the two-dimensional OU process $(y_t,z_t)^{\tr}$ above is given by 
\[
\cL_0 = -y\frac{\partial}{\partial y} - z\frac{\partial}{\partial z}
+ \sigma_N^2\frac{\partial^2}{\partial y^2} + \sigma_w^2\frac{\partial^2}{\partial z^2} .
\]
The density $p_{\infty} = \cN\bigl(0, \bigl(\begin{smallmatrix}
\sigma_N &  \\  & \sigma_w
\end{smallmatrix}\bigr)\bigr)$  
satisfies $\cL_0^{*}p_{\infty}=0$,  and is therefore the stationary density of this OU process. We wish to homogenize with respect to this density. Let $f_0:=z - (a/2)y^3$, and note that $\bbE_{p_{\infty}}f_0 = 0$ as required. The next step involves solving an appropriate Poisson problem. The process by which the Poisson PDE is derived and solved is not discussed in detail by~\cite{PardVeretI}; we direct the reader to~\citep[Chap. 11]{PavStuBook} for a more thorough treatment of the steps that follow and their justification. 
We must solve the Poisson cell problem
\[
-\cL_0\Phi(y,z) = f_0(y,z), \qquad\text{subject to }\quad \int\Phi(y,z)p_{\infty}(y,z)dydz = 0 \;.
\]
A straightforward calculation gives the centered solution
\[
\Phi(y,z) = z - a\sigma_N^2y - (a/6)y^3 .
\]
With this solution in hand, we may now compute the drift and diffusion coefficients of the approximating SDE. The drift is given by
\[
\bbE_{p_{\infty}}\bigl[-(axy^2 + (L\bx)_i)\bigr] = -(a\sigma_N^2x + (L\bx)_i)
\]
while the square of the diffusion coefficient is given by
\begin{align*}
2\bbE_{p_{\infty}}\bigl[f_0(y,z)\Phi(y,z)\bigr] &= 
2\bbE_{p_{\infty}}\bigl[\bigl(z - (a/2)y^3\bigr)\bigl(z -a\sigma_N^2y - (a/6)y^3\bigr)\bigr] \\
&= 2\sigma_w^2 + \tfrac{11}{2}a^2\sigma_N^6 \,.
\end{align*}
Putting these results together, we have that for $\varepsilon\ll 1$ and times $t$ up to $\cO(1)$ the solution to~\eqref{eqn:spsa-rescaled-sys} is approximated by the solution $\bW_t$ to
\[
d\bW_t = -(L + \gamma\sigma_N^2\alpha I)\bW_t + \sqrt{2\sigma_w^2 + \tfrac{11}{2}\gamma^2\alpha^2\sigma_N^6}\dB_t \,.
\]
We can now apply Theorem~\ref{thm:ou-sync-noisered} to this system setting $d=1, A=\gamma\sigma_N^2\alpha,\Sigma=\sqrt{2\sigma_w^2 + \tfrac{11}{2}\gamma^2\alpha^2\sigma_N^6}$, and invert the previous change of variable $\bw\to\sqrt{\varepsilon}\bw + \mu\bbone$, to obtain the estimate given in the statement of the Proposition.


\bibliographystyle{apalike}
\bibliography{syncreg_neco}

\end{document}